\documentclass[onecolumn,preprintnumbers,amsmath,amssymb,nofootinbib,showpacs]{revtex4}

\usepackage{epsf}
\usepackage{graphicx}
\usepackage{bm}
\usepackage{amsmath}

\newcommand{\be}{\begin{equation}}
\newcommand{\ee}{\end{equation}}
\newcommand{\bea}{\begin{eqnarray}}
\newcommand{\eea}{\end{eqnarray}}
\newcommand{\bdm}{\begin{displaymath}}
\newcommand{\edm}{\end{displaymath}}
\newcommand{\beas}{\begin{eqnarray*}}
\newcommand{\eeas}{\end{eqnarray*}}
\newcommand{\av}[1]{\left< #1\right>}

\newcommand{\dom}[1]{#1_{\mathcal{D}}}
\newcommand{\Hd}{\frac{\dom{\dot{a}}}{\dom{a}}}
\newcommand{\Qd}{\mathcal{Q}_{\mathcal{D}}}
\newcommand{\Rd}{\mathcal{R}_{\mathcal{D}}}
\newcommand{\Pd}{\mathcal{P}_{\mathcal{D}}}
\newcommand{\Td}{\mathcal{T}_{\mathcal{D}}}
\newcommand{\Sd}{\mathcal{S}_{\mathcal{D}}}
\newcommand{\Fd}{\mathcal{F}_{\mathcal{D}}}

\newcommand{\bkr}{\overline{\rho}}
\newcommand{\bkp}{\overline{p}}

\newcommand{\om}{\Omega_{M}}
\newcommand{\od}{\Omega_{d}}

\newcommand{\ode}{\Omega_d^e}

\begin{document}

\preprint{HD-THEP-08-30}

\title{Averaging Robertson-Walker Cosmologies}

\author{Iain A. Brown}
\email{I.Brown@thphys.uni-heidelberg.de}
\author{Georg Robbers}
\email{G.Robbers@thphys.uni-heidelberg.de}
\affiliation{Institut f\"ur Theoretische Physik, Philosophenweg 16, 69120 Heidelberg, Germany}
\author{Juliane Behrend}
\email{Juliane.Behrend@uni-ulm.de}
\affiliation{Institut f\"ur Theoretische Physik, Albert-Einstein-Allee 11, 89069 Ulm, Germany}

\date{\today}

\begin{abstract}
The cosmological backreaction arises when one directly averages the Einstein equations to recover an effective Robertson-Walker cosmology, rather than assuming a background a priori. While usually discussed in the context of dark energy, strictly speaking any cosmological model should be recovered from such a procedure. We apply the scalar spatial averaging formalism for the first time to linear Robertson-Walker universes containing matter, radiation and dark energy. The formalism employed is general and incorporates systems of multiple fluids with ease, allowing us to consider quantitatively the universe from deep radiation domination up to the present day in a natural, unified manner. Employing modified Boltzmann codes we evaluate numerically the discrepancies between the assumed and the averaged behaviour arising from the quadratic terms, finding the largest deviations for an Einstein-de Sitter universe, increasing rapidly with Hubble rate to a $0.01\%$ effect for $h=0.701$. For the $\Lambda$CDM concordance model, the backreaction is of the order of $\Omega_\mathrm{eff}^0\approx 4\times 10^{-6}$, with those for dark energy models being within a factor of two or three. The impacts at recombination are of the order of $10^{-8}$ and those in deep radiation domination asymptote to a constant value. While the effective equations of state of the backreactions in Einstein-de Sitter, concordance and quintessence models are generally dust-like, a backreaction with an equation of state $w_\mathrm{eff}<-1/3$ can be found for strongly phantom models.
\end{abstract}


\maketitle

\section{Introduction}
It has been recognised for some time that the cosmology one recovers from averaging an inhomogeneous distribution on a spacelike hypersurface does not in general coincide with a simple Friedmann-LeMa\^itre-Robertson-Walker (FLRW) prescription (see for example \cite{ShirokovFisher,Marochnik80,Marochnik80-2,Ellis84,Kasai92,Kasai93,Kasai95,Buchert95,BuchertEhlers95,Futamase96,Mukhanov96,Zalaletdinov96,Boersma97,Russ97,Unruh98,Stoeger99,Buchert99,Buchert01,Buchert02,Martineau05,Coley05,Coley06,Buchert07} for an inexhaustive list). Instead, one recovers correction terms in the averaged Friedmann and Raychaudhuri equations which when written as an effective fluid could in principle act as a dark energy. This has two clear advantages over the standard picture -- it involves no new physics, and the growth of large inhomogeneities expected to cause large deviations from the assumed behaviour coincides with the acceleration of the universe. It is with this motivation that most studies of averaged and inhomogeneous cosmologies since 1997 have been pursued (see e.g. \cite{Wetterich03,Geshnizjani02,Geshnizjani03,Rasanen03,Brandenberger04,Rasanen04,Kolb04,Kolb04-2,Geshnizjani04,Flanagan05,Kolb05,Hirata05,Geshnizjani05,Parry06,Alnes06,Rasanen06,Buchert06,Biswas06,Kasai06,Paranjape06,Ishibashi06,Tanaka06,Vanderveld06,Brouzakis06,Alnes07,Marra07,Kasai07,Biswas07,Li07,Mattsson07,Ishak07,Vanderveld07,Khosravi07,Hossain07,Wiltshire07,Wiltshire07-2,Leith07,Enqvist07,Marra07,Marra07-2,Behrend08,Enqvist08,Vanderveld08,Acoleyen08,Rosenthal08,Rasanen08,Li08,Paranjape08-1,Paranjape08,Larena08,Kolb08} and their references). Recent studies have begun to quantitatively  evaluate the ``backreaction'', either in perturbation theory (as in, for example, \cite{Li07,Li08,Behrend08,Paranjape08}) where the results are consistently of the order of $10^{-5}$ for linear modes and growing somewhat larger for smaller scales \cite{Li08}, or in approximate models of inhomogeneous universes, such as in \cite{Marra07-2,Rasanen08}, in the latter of which backreactions of the order of 0.1\% are found on scales of 100Mpc$^{-1}$.

While it is encouraging that so much attention is being focused on the averaging problem and impact of inhomogeneities in cosmology the cosmological averaging, or ``fitting'', problem is distinct from the dark energy problem. The prospect of so elegantly solving the coincidence problem is certainly alluring but the spatial averaging scheme popularised by Buchert -- or Zalaletdinov's macroscopic gravity \cite{Zalaletdinov04,Paranjape07,Paranjape08-1,Paranjape08} -- is applicable to any cosmology. The FLRW cosmology can be obtained by foliating spacetime with maximally-symmetric 3-surfaces. Since in the real, inhomogeneous universe these slices are not themselves maximally-symmetric, this involves an implicit averaging which should ideally be made explicit. Given a well-defined tensorial averaging procedure, the ``cosmological metric'' would then be that resulting from a direct averaging of the inhomogeneous metric on the 3-surfaces, and may well take the Robertson-Walker form. Since the Einstein tensor is nonlinear, however, the Einstein tensor constructed from such an averaged metric will not be the same as the averaged Einstein tensor, and the dynamics recovered from the average metric and from averaging the Einstein equations will differ. The difference between the two dynamics is known as the ``cosmological backreaction''.

Perfect fluid universes (including various scalar-field models) were considered by Buchert \cite{Buchert01} but were restricted to single-fluid models. An earlier study \cite{Russ97} used a similar formalism but evaluated the effect using multi-fluid transfer functions. Studies of late-time backreaction since have tended to focus on pure dust Einstein-de Sitter (EdS) models, although \cite{Paranjape08} included a simplified treatment of radiation alongside the dust components. In a previous study \cite{Behrend08} (hereafter BBR08), we considered a concordance $\Lambda$CDM universe with $\sim 5\%$ baryons and $\sim 25\%$ cold dark matter but our formalism was not applicable to a universe with radiative or dark energetic components. Backreaction in scalar-field universes has also been previously considered by other authors (e.g. \cite{Geshnizjani02,Geshnizjani03,Martineau05-1}) but these have focused either on inflationary modes or on super-horizon scales. In 2003, Wetterich \cite{Wetterich03} (henceforth W03) considered a universe containing a clumped cosmon field and found a deviation from FLRW behaviour potentially of order unity, with an effective equation of state $w_{\mathrm{eff}}\approx -1/15$; see also \cite{Li01} for a study of backreaction on super-horizon scales from quintessence with similarly significant results.

In the present work, we extend our previous formalism to general systems of multiple fluids and scalar fields, applicable to realistic inhomogeneous universes. Our formalism allows for the easy incorporation of arbitrary numbers of fluids in general metrics and a variety of slicings, and could be applied to a wide range of fully inhomogeneous systems in suitable coordinates. For numerical concreteness, we then consider perturbed FLRW models; the formalism we have developed makes it straightforward to consider a realistic universe containing some or all of baryonic matter, cold dark matter, radiative species, a cosmological constant, and dark energy fluids, across an arbitrary range of redshifts. This treatment includes fluid projection terms that were neglected in BBR08. We then evaluate the backreaction arising from quadratic terms and its effective equation of state with modified versions of the cmbeasy \cite{Doran03,Behrend08} and CMBFast \cite{Seljak96,Zaldarriaga97,Zaldarriaga99} Boltzmann codes, allowing quantitatively concrete evaluations for general models without the restrictions of employing, for example, approximate transfer functions as in \cite{Russ97}.

Our findings agree with and extend our previous results. The largest deviations arise in EdS universes, with a low-Hubble rate model ($h=0.45$) generating a backreaction with an effective energy density $\Omega^0_\mathrm{eff}=5\times 10^{-5}$, and a model with $h=0.701$ an effective energy density of $\Omega^0_\mathrm{eff}=1.2\times 10^{-4}$, comparable to that in radiation. The effective equation of state is $w^0_\mathrm{eff}=1/57$. In deep radiation domination, EdS models with varying Hubble rates tend to the same constant impact. The deviations from FLRW behaviour in the concordance $\Lambda$CDM model are suppressed compared to the EdS case, with $\Omega_\mathrm{eff}=4.5\times 10^{-6}$ and $w^0_\mathrm{eff}\approx 1/120$. In contrast to an EdS model, in the $\Lambda$CDM case the present-day effective equation of state depends on the Hubble rate. The impact at recombination is of the order of $10^{-8}$, potentially raising the possibility of a detection on the microwave background with a sufficiently accurate probe.

The backreaction from a dark energy model with $w_\phi=\mathrm{const}$ is similar to $\Lambda$CDM, with a slight increase in $\Omega^0_\mathrm{eff}$ due to dark energy perturbations. The effective equations of state are likewise similar, and increase monotonically with decreasing $w_\phi$ to an asymptote of $w^0_\mathrm{eff}\approx 1/57$ for strongly phantom fields. Test cases for dynamical dark energy are provided by the standard parameterisation of the equation of state $w(a)=w_0+(1-a)w_a$, an exponential potential, an inverse power-law potential and an early dark energy model. The exponential model is a tracking field and resembles EdS, but generates a smaller backreaction with $\Omega^0_\mathrm{eff}\approx 7\times 10^{-6}$ and $w^0_\mathrm{eff}\approx 1/70$. The other models are accelerating models and produce a backreaction similar to that of $\Lambda$CDM. Dark energy models generally produce a smaller backreaction than $\Lambda$CDM due to the influence of dark energy perturbations, and the effective equations of state are always greater than zero. Finally, we consider models with a non-canonical speed of sound, approximating a ``clumping'' dark energy by setting the rest-frame sound speed to zero for a range of constant equations of state $w_\phi$. Doing so, the effective equation of state peaks at $w^0_\mathrm{eff}\approx 1/30$ for $w_\phi\approx -1/2$ and declines rapidly for smaller $w_\phi$. A small but accelerating backreaction with $w^0_\mathrm{eff}<-1/3$ is found for phantom fields with $w_\phi<-1.87$. In combination with the results in W03, our study suggests that a more refined calculation is necessary to verify that the average behaviour of a universe filled with a scalar field indeed coincides with a universe filled with a homogeneous scalar field, and also that sub-horizon backreaction could act as a brake on quintessence models analogous to that in \cite{Li01}.


We begin in \S\ref{Buchert} with a brief overview of the averaging scheme and write the averaged cosmological equations for a general system of multifluids in a relatively general 3+1 split. In \S\ref{NewtGauge} we apply this formalism to a spatially-curved Robertson-Walker universe in Newtonian gauge. In \S\ref{Results} we first derive some useful approximate solutions in matter and radiation domination before numerically calculating the backreaction from quadratic terms in a range of perturbed models.
We finish with a discussion in \S\ref{Discussion}.

\section{Cosmological Averaging}
\label{Buchert}
We present only the essentials of our approach and refer readers to \cite{Buchert07,Buchert01} and BBR08 for further details. Let us foliate spacetime with 3-surfaces with vanishing shift vector and the normal to the surfaces
\be
  n^\mu=\left(\frac{1}{\alpha},\mathbf{0}\right) .
\ee
This system has the line-element
\be
  ds^2=-\alpha^2dt^2+h_{ij}dx^idx^j .
\ee
An arbitrary matter source with stress-energy tensor $T_{\mu\nu}$ has a density and stress tensor given on the hypersurface by
\be
  \varrho=n^\mu n^\nu T_{\mu\nu}, \quad S_{ij}=T_{ij} .
\ee
In a more general scheme capable of averaging tensorial quantities we would include the impacts from the current but as this approach considers only the Hamiltonian constraint and the evolution of the extrinsic curvature this is not necessary here.

Denoting an average in a spatial domain $\mathcal{D}$ with angle brackets, one may derive \cite{Russ97,Buchert99,Buchert01,Behrend08} from the Hamiltonian constraint and the evolution of the extrinsic curvature an averaged ``Friedmann'' equation and an averaged ``Raychaudhuri'' equation,
\be
\label{BuchertEqq1}
  \left(\Hd\right)^2=\frac{8\pi G}{3}\av{\alpha^2\varrho}+\av{\alpha^2}\frac{\Lambda}{3}-\frac{1}{6}\left(\Qd+\Rd\right), \quad
  \frac{\dom{\ddot{a}}}{\dom{a}}=-\frac{4\pi G}{3}\av{\alpha^2\left(\varrho+S\right)}+\av{\alpha^2}\frac{\Lambda}{3}+\frac{1}{3}\left(\Qd+\Pd\right) .
\ee
These are often known as the Buchert equations, although it should be noted that they have been rederived multiple times with different notations (e.g. \cite{Marochnik80,Russ97}). The kinematic and dynamic backreactions and the curvature correction are
\be
  \Qd=\av{\alpha^2\left(K^2-K^i_jK^j_i\right)}-\frac{2}{3}\av{\alpha K}^2, \quad
  \Pd=\av{\alpha D^iD_i\alpha}-\av{\dot{\alpha}K}, \quad
  \Rd=\av{\alpha^2\mathcal{R}}
\ee
where $D_i$ is the covariant derivative on the 3-surface, $\mathcal{R}$ the Ricci scalar on the 3-surface, and
\be
  K_{ij}=-\frac{1}{2\alpha}\dot{h}_{ij}
\ee
the extrinsic curvature. The kinematic backreaction is related to the variance of the extrinsic curvature, the dynamic backreaction vanishes in a synchronous slicing, and the curvature correction is a direct average of the Ricci 3-scalar. The Hubble rate is defined by the volume expansion as
\be
  \Hd=\frac{1}{3}\frac{\dot{V}}{V}=-\frac{1}{3}\av{\alpha K}.
\ee
From these equations (\ref{BuchertEqq1}) one may further derive an integrability condition connecting the backreaction and curvature terms. In a general slicing and with general fluids its form is rather complicated (it is presented in BBR08), although it simplifies significantly for a synchronous slicing when the dynamical backreaction vanishes, and more so for dust models (see \cite{Buchert01} for details) for which it is of significant use \cite{Li07,Li08}.

We consider a universe filled with baryons, cold dark matter, photons, three species of massless neutrinos, a minimally-coupled scalar field, and a cosmological constant. The perfect fluid provides us with a good description of all of these. The underlying model treats radiation in the usual manner as effective fluids governed by Boltzmann hierarchies and so no physical error is introduced in doing so; the averages are purely formal and we can unambiguously map the lower moments of the hierarchies onto the fluid density and velocity. A perfect fluid with 4-velocity
\be
\label{Perfect4Vel}
  u^\mu=\frac{1}{\alpha}\left(\sqrt{1+h_{ij}v^iv^j},\alpha v^i\right)
\ee
normalised to $u^\mu u_\mu=-1$, rest-frame density $\rho$ and rest-frame isotropic pressure $p$ has the stress-energy tensor
\be
  T_{\mu\nu}=\rho u_\mu u_\nu+p\tilde{P}_{\mu\nu}, \quad \tilde{P}_{\mu\nu}=g_{\mu\nu}+u_\mu u_\nu .
\ee
The projection tensor $\tilde{P}_{\mu\nu}$ projects quantities into the fluid rest-frame. The forms for baryons and CDM are then recovered by setting $p=0$, those for radiative fluids by setting $p=\rho/3$ and that for a scalar field with $p=w_\phi\rho$. The fluid contributions to the average cosmological equations for a species $(a)$ can then be written
\be
  \frac{8\pi G}{3}\av{\alpha^2\varrho_{(a)}}=\frac{8\pi G}{3}\av{\alpha^2\rho_{(a)}}+\Fd^{(a)}, \quad
  -\frac{4\pi G}{3}\av{\alpha^2S_{(a)}}=-4\pi G\av{\alpha^2p}-\frac{1}{2}\Fd^{(a)}
\ee
where
\be
  \Fd^{(a)}=\frac{8\pi G}{3}\av{\alpha^2\left(n^\mu n^\nu-u^\mu_{(a)} u^\nu_{(a)}\right)T^{(a)}_{\mu\nu}}=
     \frac{8\pi G}{3}\av{\alpha^2\left(h^{\mu\nu}-\tilde{P}^{\mu\nu}_{(a)}\right)T^{(a)}_{\mu\nu}}
\ee
accounts for the tilt between the fluid rest-frames and the foliation.

The multifluid averaged cosmology can then be written as
\bea
  \left(\Hd\right)^2&=&\frac{8\pi G}{3}\sum_a\av{\alpha^2\rho_{(a)}}+\frac{1}{3}\av{\alpha^2}\Lambda-\frac{1}{6}\left(\Rd+\Qd-6\sum_a\Fd^{(a)}\right), \\
  \frac{\dom{\ddot{a}}}{\dom{a}}&=&-\sum_a\frac{4\pi G}{3}\av{\alpha^2\left(\rho_{(a)}+3p_{(a)}\right)}+\frac{1}{3}\av{\alpha^2}\Lambda+\frac{1}{3}\left(\Pd+\Qd-3\sum_a\Fd^{(a)}\right) .
\eea
It is tempting to immediately define an effective energy density, pressure and equation of state of the ``backreaction fluid''
\bea
\label{EffDensity}
  \frac{8\pi G}{3}\overline{\rho}_{\mathrm{eff}}&=&\sum_a\Fd^{(a)}-\frac{1}{6}\left(\Qd+\Rd\right), \\
\label{EffPressure}
  16\pi G\overline{p}_{\mathrm{eff}}&=&\frac{16\pi G}{3}\overline{S}_{\mathrm{eff}}=2\sum_a\Fd^{(a)}+\frac{1}{3}\left(\Rd-3\Qd-4\Pd\right)
\eea
and so
\be
\label{EffEquationOfState}
  w_{\mathrm{eff}}=-\frac{1}{3}\frac{\Rd-3\Qd-4\Pd+6\sum_a\Fd^{(a)}}{\Rd+\Qd-6\sum_a\Fd^{(a)}} .
\ee
(Compare with, for example, equation (45) of BBR08). Note, however, that while this does give a correction to an effective FLRW it does not let us directly compare with an \emph{input} FLRW model: $\Rd$ contains a background curvature and, depending on the choice of $\alpha$, $\Pd$ could contain background functions of the Hubble rate. Furthermore, additional non-linear contributions arise from the fluid and cosmological constant terms. Accordingly, the effective fluid is defined in a model-specific manner.

\section{Newtonian Gauge Perturbation Theory}
\label{NewtGauge}
The system of equations presented in \S\ref{Buchert} is general for any system for which the 3+1 split remains valid. To make a quantitative evaluation of the impact of the backreaction, we must specify a system. The simplest first approximation is a perturbed Robertson-Walker metric, which as in BBR08 we consider in Newtonian gauge. The use of perturbations and this gauge is subject to the same caveats as before. Firstly, linear (or mildly non-linear) theory restricts us to averages on large scales, which in general will be outwith our past light-cone and not necessarily observable. Secondly, using perturbation theory automatically assumes that the backreaction terms are small -- otherwise, the background would be poorly chosen. While this is undoubtedly so, and while it cannot na\"ively be taken to give us the impact observed on the CMB (although \cite{Larena08} attempts to connect a similar formalism with CMB observables) or in supernovae surveys for which averages across much smaller scales (such as those in \cite{Li08,Rasanen08,Buchert08}) must be performed, the use of perturbation theory means that we can quantitatively evaluate backreaction terms in the simplest realistic cases, which can provide us with some of the features of a more general model. Moreover, linear theory is still valid on scales above $k\approx 0.06h\mathrm{Mpc}^{-1}$, as seen from the matter power spectrum \cite{Percival06}, and at the epoch of recombination and in radiation domination the universe is extremely well described by perturbation theory. Second-order theory, even in the present epoch, is valid to slightly smaller scales. In these cases our calculations will be robust.

It is also worth commenting on the different scale factors we employ. When we use a perturbed Robertson-Walker universe as the underlying model the scale factor $a(t)$ is that associated with the metric, appearing in and governed by the Einstein equations. The scale factor $a_\mathcal{D}(t)$ appearing in the averaged cosmological equations is, in contrast, a ``reconstructed'' scale factor, defined by an average of an inhomogeneous distribution across the domain $\mathcal{D}$. Its evolution is governed by the averaged equations and clearly has no impact on the underlying model. The distinction between the two is extremely important.

This issue arises because of the somewhat circular nature of the argument: we are assuming a large-scale ``average'', adding perturbations to approximate the inhomogeneous universe and re-averaging the results. Phrased differently, we are defining perturbations on a smooth ``background'' manifold, and then averaging across the inhomogeneous manifold defined by the perturbations. In a more realistic model, one would not be employing an underlying Robertson-Walker model at all but rather a distribution of sources (as in \cite{Rasanen08}) or multi-scale models (as in \cite{Wiltshire07,Buchert08}), and the only scale factor would be the reconstructed $\dom{a}$, and this confusion would not arise.

Take the Newtonian metric in the form
\be
\label{LineElement}
  ds^2=-(1+2\Psi)dt^2+a^2(t)(1-2\Phi)\gamma_{ij}dx^idx^j
\ee
where the spatial metric is
\be
  \gamma_{ij}=\mathrm{diag}\left(\frac{1}{1-\mathcal{K}r^2}, r^2, r^2\sin^2(\theta)\right)
\ee
and we work with spherical polar co-ordinates $x^i=(r,\theta,\phi)$. The perturbations $\Psi$ and $\Phi$ are non-linear and can if required be expanded in a Taylor series; we restrict ourselves to second-order, retaining up to quadratic terms in perturbations. Therefore
\be
  \Phi=\sum_n\frac{1}{n!}\Phi_{(n)}=\Phi_{(1)}+\frac{1}{2}\Phi_{(2)}, \quad \Psi=\sum_n\frac{1}{n!}\Psi_{(n)}=\Psi_{(1)}+\frac{1}{2}\Psi_{(2)}, \quad \Phi\Psi=\Phi_{(1)}\Psi_{(1)}
\ee
with similar results for $\Phi^2$ and $\Psi^2$. We then immediately have
\be
  \alpha^2=1+2\Psi, \quad \alpha=1+\Psi-\frac{1}{2}\Psi^2+\mathcal{O}(\Psi^3), \quad \alpha^{-1}=1-\Psi+\frac{3}{2}\Psi^2+\mathcal{O}(\Psi^3), \quad \alpha^{-2}=1-2\Psi+4\Psi^2+\mathcal{O}(\Psi^3)
\ee
and
\be
  h_{ij}=a^2(t)(1-2\Phi)\gamma_{ij}, \quad h^{ij}=a^{-2}(t)(1+2\Phi+4\Phi^2)\gamma^{ij}+\mathcal{O}(\Phi^3) .
\ee
Following BBR08, the extrinsic curvature is
\be
  \alpha K^i_j=-\left(\frac{\dot{a}}{a}-(1+2\Phi)\dot{\Phi}\right)\delta^i_j
\ee
and so the kinematical backreaction is
\be
  \Qd=6\left(\av{\dot{\Phi}^2}-\av{\dot{\Phi}}^2\right),
\ee
trivially related to the variance of $\dot{\Phi}$. Expanding the covariant derivative on the Newtonian slicing and using the time derivative of the lapse, we find the dynamical backreaction
\bea
  \Pd&=&\frac{1}{a^2}\av{\nabla^2\Psi-(\nabla\Psi)^2+2\Phi\nabla^2\Psi-(\nabla\Phi)\cdot(\nabla\Psi)}
     +3\frac{\dot{a}}{a}\av{\dot{\Psi}-2\Psi\dot{\Psi}}-3\av{\dot{\Psi}\dot{\Phi}}
   \nonumber \\ && \quad
\label{PdDef}
     -\frac{\mathcal{K}}{a^2}\av{3r\left(\frac{\partial\Psi}{\partial r}+2\Phi\frac{\partial\Psi}{\partial r}\right)+r^2\left(\frac{\partial^2\Psi}{\partial r^2}+2\Phi\frac{\partial^2\Psi}{\partial r^2}+\left(1+\frac{\partial\Phi}{\partial r}\right)\frac{\partial\Psi}{\partial r}\right)}
\eea
where $\nabla=(\partial/\partial r, (1/r)\partial/\partial\theta, (1/r\sin\theta)\partial/\partial\phi)$ is the usual Euclidean gradient operator and the curvature corrections are written explicitly. The Ricci scalar on the 3-surfaces generates the curvature term
\bea
  \Rd&=&\frac{6\mathcal{K}}{a^2}+\frac{2}{a^2}\av{2\nabla^2\Phi+3(\nabla\Phi)^2+4(2\Phi+\Psi)\nabla^2\Phi}
    +\frac{4\mathcal{K}}{a^2}\av{\left(\Phi+\Psi\right)-3r\frac{\partial\Phi}{\partial r}-r^2\frac{\partial^2\Phi}{\partial r^2}}
  \nonumber \\ && \quad
\label{RdDef}
    -\frac{2\mathcal{K}}{a^2}\av{4\left(\Psi+2\Phi\right)r^2\frac{\partial^2\Phi}{\partial r^2}+3r^2\left(\frac{\partial\Phi}{\partial r}\right)^2+12\left(\Psi+2\Phi\right)r\frac{\partial\Phi}{\partial r}-12\Phi\left(\Psi+\Phi\right)} .
\eea
As expected, this contains the background curvature from the input FLRW model.

The fluid 4-velocity is
\be
  u^\mu=\left(1-\Psi+\frac{1}{2}\left(a^2\mathcal{V}^2+3\Psi^2\right),v_r,v_\theta,v_\phi\right)
\ee
and its norm is
\be
  h_{ij}v^iv^j=a^2\mathcal{V}^2=a^2\left(\frac{v_r^2}{1-\mathcal{K}r^2}+r^2v_\theta^2+r^2\sin^2\theta v_\phi^2\right).
\ee
We linearising the density and the pressure with respect to the FLRW background,
\be
  \rho=\overline{\rho}\left(1+\delta\right), \quad
  p=\overline{p}\left(1+\gamma\right),
\ee
where again $\delta$ and $\gamma$ contain contributions from first- and second-order. This then gives the total fluid contributions from this species as
\be
  \frac{8\pi G}{3}\av{\alpha^2\rho}+\Fd=\frac{8\pi G}{3}\bkr+\Td, \quad
  \frac{4\pi G}{3}\av{\alpha^2(\rho+S)}+\Fd=\frac{4\pi G}{3}\left(\bkr+3p\right)+\frac{1}{2}\Td+\Sd
\ee
where the corrections to the FLRW case have been consolidated into a density correction $\Td$ and a pressure correction $\Sd$, defined by
\be
\label{TdSdDef}
  \frac{3\Td}{8\pi G}=\overline{\rho}\av{\delta+2\Psi+a^2(1+\overline{w})\mathcal{V}^2+2\Psi\delta}, \quad
  \frac{3\Sd}{4\pi G}=\overline{\rho}\av{3c_s^2\delta+6\overline{w}\Psi+a^2(1+\overline{w})\mathcal{V}^2+6c_s^2\Psi\delta} .
\ee
Setting $\mathcal{K}=0$, $w=c_s^2=0$, neglecting anisotropic stresses (implying $\Phi=\Psi$) and swapping to Cartesian co-ordinates reduces these results to those in BBR08.

The effective fluid that arises in an FLRW universe perturbed to second order is therefore given by
\bea
  \frac{8\pi G}{3}\bkr_{\mathrm{eff}}&=&\Td+\frac{1}{3}\Lambda\av{\alpha^2-1}-\frac{1}{6}\left(\Qd+\Rd\right), \\
  16\pi G\bkp_\mathrm{eff}&=&4\Sd-2\av{\alpha^2-1}\Lambda+\frac{1}{3}\left(\Rd-3\Qd-4\Pd\right)
\eea
where it is understood that $\Rd$ does not contain the background term. We will characterise the ``backreaction'' with the redundant set of dimensionless quantities, $\left\{\Omega_\mathrm{eff},\Delta R/R,w_\mathrm{eff}\right\}$ where
\bea
\label{OmegaEff}
  \Omega_{\mathrm{eff}}&=&\frac{8\pi G\bkr_{\mathrm{eff}}}{3(\dot{a}/a)^2}, \\
\label{DeltaRoR}
  \frac{\Delta R}{R}&=&-\frac{4\pi G\left(\bkr_\mathrm{eff}+3\bkp_\mathrm{eff}\right)}{\left|\ddot{a}/a\right|}=\frac{(1/3)\left(\Qd+\Pd\right)-\sum_a\left(\frac{1}{2}\Td^{(a)}+\Sd^{(a)}\right)}{\left|\ddot{a}/a\right|}, \\
\label{EquationOfState}
  w_\mathrm{eff}&=&\frac{\bkr_\mathrm{eff}}{\bkp_\mathrm{eff}}=-\frac{1}{3}\frac{\Rd-4\Pd-3\Qd+12\Sd-12\Lambda\av{\alpha^2-1}}{\Rd+\Qd-6\Td-4\Lambda\av{\alpha^2-1}} .
\eea
$\Omega_\mathrm{eff}$ and $\Delta R/R$ give the corrections to the Friedman and Raychaudhuri equations respectively that an observer in the perturbed FLRW universe would reconstruct. Normalising the correction to the Raychaudhuri equation in this manner introduces an artificial singularity if the FLRW model passes from deceleration to acceleration (or vice-versa); if desired this can be avoided by normalising instead to the contribution from, for example, dust matter.

\section{Results from Quadratic Modes}
\label{Results}
When averaging across very large scales, due to their Gaussian nature it is a reasonable approximation to neglect averages of pure first-order perturbations. The same argument cannot be applied to second-order perturbations. However, it is not clear how in general an average of the form $\av{\delta_{(2)}}$ can be performed, although analytic solutions could be employed in particular, highly simplified, cases. Here we consider solely the impact from the quadratic terms and leave the general issue to future study. While a calculation from purely quadratic terms is strictly neither a complete nor a consistent calculation, it does give a firm order-of-magnitude estimate of the effect. Since second-order quantities can be written using the Einstein equations as combinations of first-order quantities, the error introduced is at most approximately of order $\mathcal{O}(1)$ and calculations from the quadratic terms treated loosely as upper bounds on the impact.

If we neglect the averages of pure first-order perturbations and their derivatives, which is a reasonable assumption when averaging across very large scales, the backreactions and curvature correction reduce to
\bea
  \Qd&=&6\av{\dot{\Phi}^2}, \\
  \Pd&=&\frac{1}{a^2}\av{2\Phi\nabla^2\Psi-(\nabla\Psi)^2-(\nabla\Phi)\cdot(\nabla\Psi)}
    -6\frac{\dot{a}}{a}\av{\Psi\dot{\Psi}}-3\av{\dot{\Psi}\dot{\Phi}}
   \nonumber \\ && \quad
    -\frac{\mathcal{K}}{a^2}\av{6r\Phi\frac{\partial\Psi}{\partial r}+r^2\left(2\Phi\frac{\partial^2\Psi}{\partial r^2}+\frac{\partial\Phi}{\partial r}\frac{\partial\Psi}{\partial r}\right)}
\label{Pd}
    , \\
  \Rd&=&\frac{6\mathcal{K}}{a^2}+\frac{2}{a^2}\av{3(\nabla\Phi)^2+4(2\Phi+\Psi)\nabla^2\Phi}
   \nonumber \\ && \quad
    -\frac{2\mathcal{K}}{a^2}\av{4\left(\Psi+2\Phi\right)\left(3r\frac{\partial\Phi}{\partial r}+r^2\frac{\partial^2\Phi}{\partial r^2}\right)
    +3r^2\left(\frac{\partial\Phi}{\partial r}\right)^2-12\Phi\left(\Psi+\Phi\right)}
\label{Rd}
\eea
and the fluid corrections to
\bea
\label{Td}
  \Td^{(a)}&=&\frac{8\pi G\overline{\rho}_{(a)}}{3}\av{(1+\overline{w}_{(a)})a^2\mathcal{V}_{(a)}^2+2\Psi\delta_{(a)}}, \\
\label{Sd}
  \Sd^{(a)}&=&\frac{4\pi G\overline{\rho}_{(a)}}{3}\av{(1+\overline{w}_{(a)})a^2\mathcal{V}_{(a)}^2+6c_{s(a)}^2\Psi\delta_{(a)}}
	    =\frac{1}{2}\Td^{(a)}+\frac{8\pi G\overline{\rho}_{(a)}}{3}\left(3c_{s(a)}^2-1\right)\av{\Psi\delta_{(a)}} .
\eea
On very large scales where we can invoke the ergodic principle, we can turn the volume averages into ensemble averages. If $\mathcal{P}_\psi(k)$ is the primordial power spectrum of $\Psi$, then with $\mathcal{K}=0$ the various correction and backreaction terms become one-dimensional integrals across the primordial power spectrum:
\bea
\label{IntegralsStart}
  \Qd&=&6\int\mathcal{P}_\psi(k)\left|\dot{\Phi}\right|^2\frac{dk}{k}, \\
  \Pd&=&-3\int\mathcal{P}_\psi(k)\left(\frac{k^2}{a^2}\left(\frac{1}{2}\Phi\Psi^*+\frac{1}{2}\Phi^*\Psi+\frac{1}{3}\left|\Psi\right|^2\right)
    +\frac{\dot{a}}{a}\left(\Psi\dot{\Psi}^*+\Psi^*\dot{\Psi}\right)+\frac{1}{2}\left(\dot{\Psi}\dot{\Phi}^*+\dot{\Psi}^*\dot{\Phi}\right)
    \right)\frac{dk}{k}, \\
  \Rd&=&-\frac{2}{a^2}\int k\mathcal{P}_\psi(k)\left(5\left|\Phi\right|^2+2\Psi\Phi^*+2\Psi^*\Phi\right)dk,\\
  \Td^{(a)}&=&\frac{8\pi G}{3}\overline{\rho}_{(a)}\int\mathcal{P}_\psi(k)\left(\Psi\delta_{(a)}^*+\Psi^*\delta_{(a)}+(1+w_{(a)})a^2\left|v_{(a)}\right|^2\right)\frac{dk}{k},\\
  \Sd^{(a)}&=&\frac{4\pi G}{3}\overline{\rho}_{(a)}\int\mathcal{P}_\psi(k)\left(3c^2_{s(a)}\Psi\delta_{(a)}^*+3c^2_{s(a)}\Psi^*\delta_{(a)}+(1+w_{(a)})a^2\left|v_{(a)}\right|^2\right)\frac{dk}{k} .
\label{IntegralsEnd}
\eea
The power spectrum is defined in the usual manner as
\be
  \av{\Psi(\mathbf{k})\Psi^*(\mathbf{k}')}=\frac{2\pi^2}{k^3}\mathcal{P}_\psi(k)\delta(\mathbf{k-k}')
\ee
and we assume a power law form
\be
  \mathcal{P}_\psi(k)=A_s\left(\frac{k}{k_*}\right)^{n_s-1} .
\ee
Unless otherwise mentioned we assume the WMAPV parameters $A_s=2.4\times 10^{-9}$ and $n_s=0.96$ at a pivot $k_*=0.002\mathrm{Mpc}^{-1}$ \cite{Komatsu08}. Due to the form of the integrals, modifying the amplitude and spectral index will amount to a renormalisation of the backreaction. In principle, the expressions (\ref{IntegralsStart}-\ref{IntegralsEnd}) are not accurate; this is because the Fourier modes are defined with respect to the FLRW background, while the averages are defined with respect to the inhomogeneous background. However, the correction term can be shown to be negligible for the models we consider.

The curvature terms in equations (\ref{Pd}) and (\ref{Rd}) are non-trivial. Our approach involves computing the perturbations as a function of wavenumber $k$ and converting volume averages on large scales into ensemble averages. Doing so, we can convert gradients into powers of $ik$. However, the curvature modifications involve derivatives with respect to a radial coordinate $r$ and we cannot deal with this so simply. Rather than pursue our current approach in the curved case, we consider only flat models with $\mathcal{K}=0$. Employing a somewhat different approach, Rosenthal and Flanagan, however, recently considered the backreaction in certain closed models \cite{Rosenthal08}. In forthcoming work we will consider the spatially-curved cases in our formalism, employing an alternative direct spatial averaging approach.

\subsubsection{Analytic Approximations}
We can derive analytic solutions to the above equations in particular regimes, the most interesting of which are a universe filled with a single radiation fluid, which serves as an approximation for the real universe in deep radiation domination, and a universe filled with dust, corresponding to an EdS universe or the $\Lambda$CDM model for $z\gtrsim 1-2$. In both cases the anisotropic stress vanishes and the curvature correction and backreactions become
\be
  \Rd=-\frac{18}{a^2}\int k\mathcal{P}_\psi(k)\left|\Phi\right|^2dk, \quad
  \Qd=\frac{6}{a^2}\int\mathcal{P}_\psi(k)\left|\Phi'\right|^2\frac{dk}{k}, \quad
  \Pd=\frac{2}{9}\Rd-\frac{1}{2}\Qd-\frac{6}{a}\frac{\dot{a}}{a}\int\mathcal{P}_\psi(k)\mathrm{Re}\left(\Phi\Phi'\right)\frac{dk}{k} .
\ee
Here $'=\partial/\partial\eta$ and $\eta$ is the conformal time $d\eta=dt/a$. For a single fluid with $v^2=\mathcal{V}^2=v_x^2+v_y^2+v_z^2$,
\be
  \Td=\frac{8\pi G\bkr}{3}\int\mathcal{P}_\psi(k)\left(2\mathrm{Re}\left(\Phi\delta^*\right)+(1+w)a^2v^2\right)\frac{dk}{k}, \quad
  \Sd=\frac{1}{2}\Td+\frac{8\pi G\bkr}{3}\left(3c_s^2-1\right)\int\mathcal{P}_\psi(k)\mathrm{Re}\left(\Phi\delta^*\right)\frac{dk}{k} .
\ee
The evolution of the quantities $\Phi$, $\delta$ and $v$ is given by the Einstein and conservation equations. See for example \cite{Mukhanov92,Durrer04,Malik08} for these; here we will present only the results.

For a universe filled with a radiative fluid where $w=c_s^2=1/3$ the background evolves with respect to conformal time as
\be
  \frac{a'}{a}=\frac{1}{\eta}, \quad \frac{a''}{a}=0, \quad \frac{a}{a_\mathrm{eq}}=\frac{\eta}{\eta_\mathrm{eq}} ,
\ee
and the pressure correction becomes
\be
  \Sd=\frac{1}{2}\Td .
\ee
The perturbations can be seen (\cite{Durrer04}) to evolve on sub-horizon scales ($k\eta\gg 1$) as
\be
\label{PertsRadDom}
  av(k,\eta)=\frac{\sqrt{3}}{2}D\sin(k\eta), \quad \delta(k,\eta)=2D\cos(k\eta), \quad \Phi(k,\eta)=-\frac{D\cos(k\eta)}{k^2\eta^2} .
\ee
The matter terms are dominant and taking $n_s=1$ for simplicity implies
\be
\label{RadTd}
  \Td=2\Sd=\frac{H_0^2\Omega^0_RD^2}{a^4}\int_{x_\mathrm{min}}^{x_\mathrm{max}}\frac{\sin^2x}{x}dx
\ee
where $x=k\eta$. The  effective energy density (\ref{OmegaEff}), modification to the Raychaudhuri equations (\ref{DeltaRoR}) and effective equation of state (\ref{EquationOfState}) are then
\be
  \Omega_\mathrm{eff}\approx a^2\eta^2\Td,
  \quad \frac{\Delta R}{R}\approx -\Omega_\mathrm{eff},
  \quad w_{\mathrm{eff}}\approx\frac{1}{3} .
\ee
Setting $\eta_\mathrm{eq}$ from the present day using (\ref{MatterA}), the effective energy density is
\be
\label{DFoFRad}
  \Omega_\mathrm{eff}\approx 4D^2\Omega^0_M\int\frac{\sin^2x}{x}dx\sim\mathrm{const} .
\ee
The modifications in deep radiation domination, then, tend to a constant and have no dependence on the Hubble rate. They will, however, have a dependence on the scalar spectral index. We can also see from (\ref{DFoFRad}) that, perhaps contrary to expectation, the oscillations present in the pre-recombination plasma and in radiation domination are not expected to have a significant impact on the results.

The opposite r\'egime of interest is a universe filled with dust. Setting $w=c_s^2=0$ gives
\be
  \Sd=\frac{4\pi G\overline{\rho}}{3}\av{a^2v_{(a)}^2}\geq 0 ,
\ee
which corresponds to the ``gravitational pressure'' evaluated in W03 and BBR08. This term was otherwise neglected in BBR08, a good approximation when considering the impact on the Friedmann and Raychaudhuri equations but less so when considering the effective equation of state. In a dust universe, the perturbations evolve as
\be
  \Phi(k,\eta)=\Psi(k,\eta)=\Phi_1(k), \quad \Phi'(k,\eta)=0, \quad \delta(k,\eta)=-\frac{1}{6}\Phi_1(k)k^2\eta^2, \quad av(k,\eta)=\frac{1}{3}\Phi_1(k)k\eta
\ee
with the background
\be
  \frac{a'}{a}=\frac{2}{\eta}, \quad \frac{a''}{a}=\frac{2}{\eta^2}, \quad a=\frac{\eta^2}{\eta_0^2} .
\ee
The correction terms are then
\be
  \Qd=0, \quad \Pd=\frac{2}{9}\Rd, \quad \Rd=-\frac{18}{a^2}\int k\mathcal{P}_\psi(k)\Phi_1^2(\mathbf{k})dk, \quad
  \Td=\frac{4}{81}\Rd, \quad \Sd=-\frac{1}{4}\Td .
\ee
These exact results agree with the purely numerical results from BBR08, where $\Rd/\Pd\approx 4.5$ and $\Td/\Rd\approx 1/20$. The effective equation of state is
\be
  w_\mathrm{eff}=\frac{1}{57}
\ee
The gravitational pressure thus makes the effective equation of state from linear perturbations greater than zero -- neglecting $\Sd$ recovers $w_\mathrm{eff}=-1/19$, as found before. The effective energy density and change to the Raychaudhuri equation are
\be
\label{EdSDeltaFoF}
  \Omega_\mathrm{eff}\approx \frac{19}{9H_0^2}a\int_{k_\mathrm{min}}^{k_\mathrm{max}}k\mathcal{P}_\psi(k)\Phi_1^2(k)dk, \quad
  \frac{\Delta R}{R}\approx-\frac{20}{19}\Omega_\mathrm{eff} .
\ee
The backreaction in matter domination thus evolves linearly with the scale factor. Since the impact at the present day is expected to be of the order of $\Omega^0_\mathrm{eff}\approx 10^{-5}$, the backreaction at last scattering is $\Omega_\mathrm{eff}\approx 10^{-8}$. Our calculations, arising from linear theory, can only be interpreted at the present epoch as an order of magnitude estimate of the backreaction from large-scale modes, and the issue of backreaction from highly inhomogeneous structure is necessarily left unaddressed; at the last scattering surface, however, the universe is well-described by linear perturbation theory and our result is robust. The level of deviation from Robertson-Walker behaviour at recombination is then similar in magnitude to the standard polarisation anisotropies, suggesting it may be observable with a suitably sensitive probe and a method of mapping the spatially-averaged cosmology onto observable quantities. The question of impacts from backreaction on the CMB was recently discussed in \cite{Larena08} using scaling assumptions for the correction terms.

As in the radiative case, a change in the primordial power spectrum will merely rescale the modifications by a constant factor. However, in contrast, they are sensitively dependent on the Hubble rate. While (\ref{EdSDeltaFoF}) might seem to imply that a higher Hubble rate will induce a lower backreaction, this is not the case. $\Phi_1^2(k)$ is set at the epoch of matter-radiation equality and by (\ref{PertsRadDom}) on small scales is approximately of the form
\be
  \Phi_1(k)\approx -A\frac{\cos(k\eta_\mathrm{eq})}{k^2\eta_\mathrm{eq}^2}
\ee
where
\be
\label{MatterA}
  \eta^2_\mathrm{eq}\approx a_\mathrm{eq}\eta_0^2\approx 4\frac{a_\mathrm{eq}}{H_0^2} .
\ee
This scales (\ref{EdSDeltaFoF}) by a factor of $H_0^4$. Furthermore, we choose to take as our domain the Hubble volume, and so $k_\mathrm{min}\approx\pi H_0$ at the present epoch and the limit introduces further factors of $H_0$. We characterise the result by
\be
  \Omega^0_\mathrm{eff}\propto h^{2-m}
\ee
where $H_0=h\times 100\mathrm{kms}^{-1}\mathrm{Mpc}^{-1}$ and we expect $m$ to be small. The dependence on the Hubble rate is then not necessarily trivial, but we expect an increase in the Hubble rate to increase the signature from backreaction.

\subsubsection{Einstein-de Sitter and $\Lambda$CDM}
In BBR08 we considered numerically an EdS and a $\Lambda$CDM universe for $z\lesssim 100$, finding $w_\mathrm{eff}\approx -1/19$ at the present day, declining slightly as $z\rightarrow 100$. However, we neglected the ``gravitational pressure'' term $\Sd$ and were constrained to low redshifts and pure dust models. Here we update the treatment to the WMAPV concordance cosmology \cite{Dunkley08,Komatsu08}, high redshifts and dark energy cosmologies. We use modified versions of the cmbeasy \cite{Doran03} and CMBFast \cite{Seljak96} Boltzmann codes (collectively referred to as Backfast) and numerically integrate equations (\ref{IntegralsStart}-\ref{IntegralsEnd}) as a function of time. Our domain is $k\in(2\pi/\eta,40\mathrm{Mpc}^{-1})$ where the small-scale limit is set such that the integrals converge. Naturally we do not claim that perturbation theory actually applies on such scales, merely that our results give the total contribution from such modes.

\begin{center}
\begin{figure}
\includegraphics[width=0.45\textwidth]{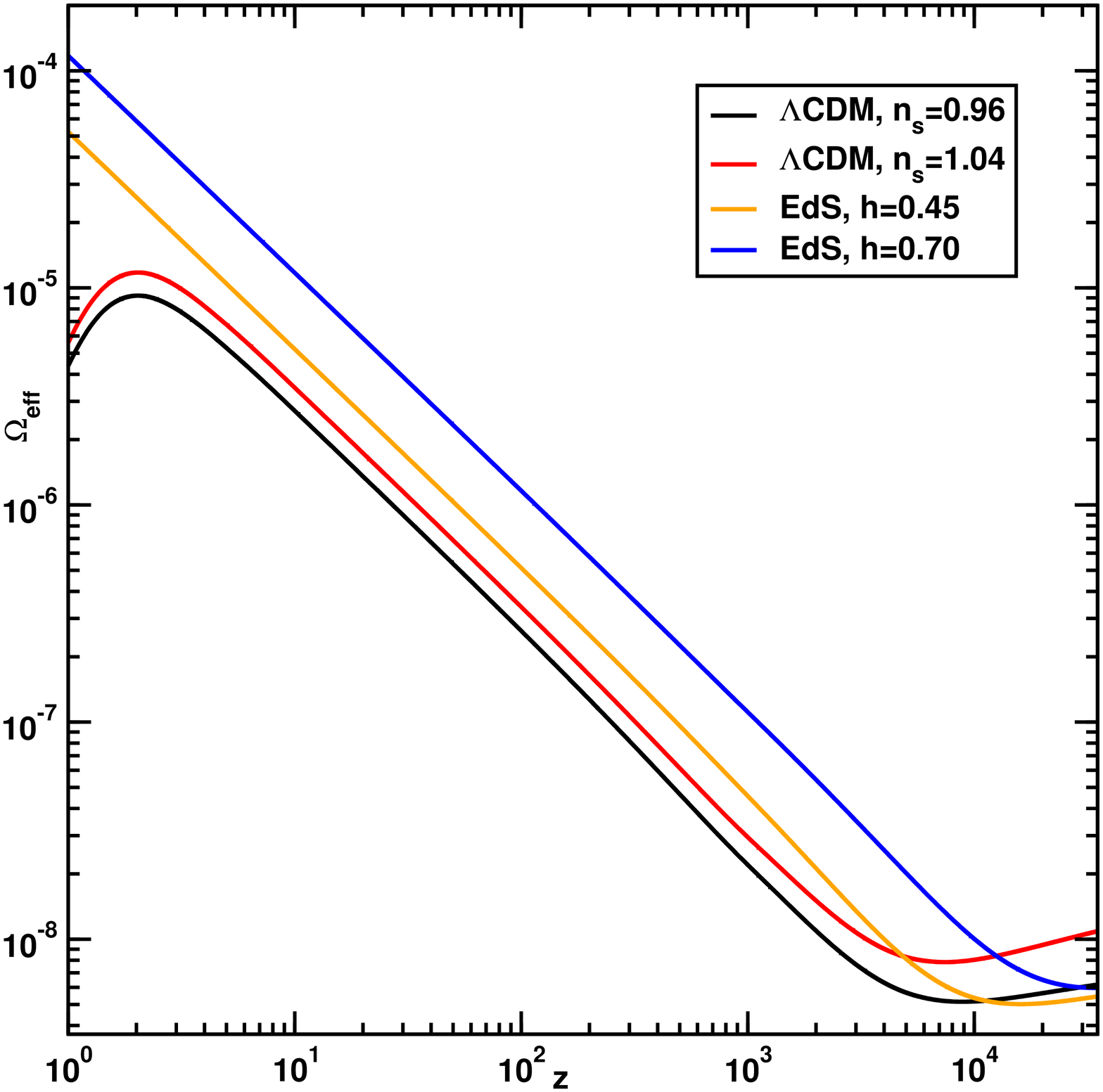}\;
\includegraphics[width=0.45\textwidth]{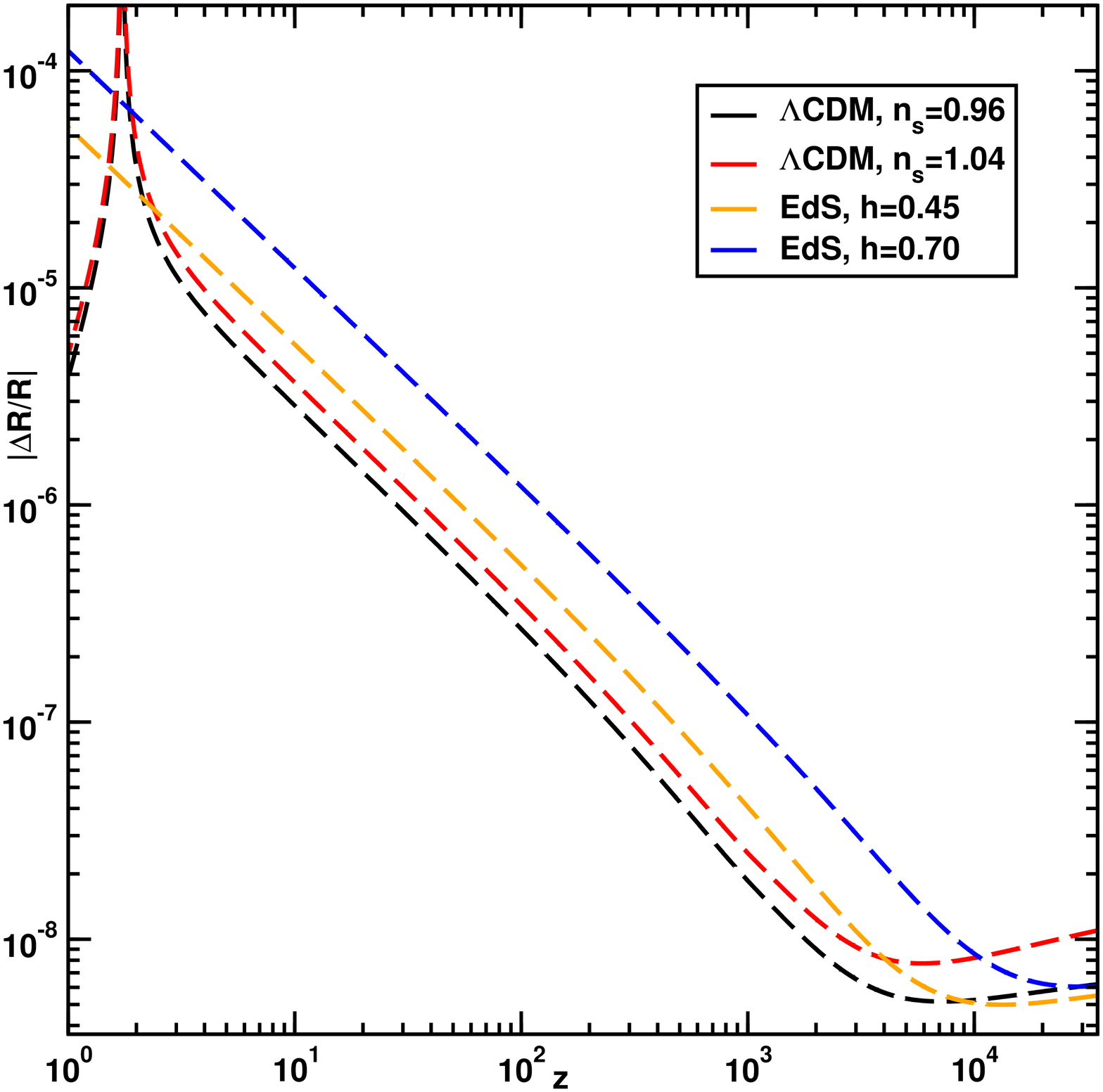}
\caption{Effective energy density of backreaction and modification to the Raychaudhuri equation for EdS and $\Lambda$CDM. Dashed lines are negative. Left: $\Omega_\mathrm{eff}$ for $\Lambda$CDM with $n_s=0.96$ and $n_s=1.04$, and an EdS model with $h=0.45$ and $h=0.701$. Right: $-\Delta R/R$ for the same models at high redshift. The impacts are in agreement with the approximate predictions.}
\label{FriedRay-LCDMEdS}
\end{figure}
\end{center}

Figure \ref{FriedRay-LCDMEdS} shows $\Omega_\mathrm{eff}$ and $|\Delta R/R|$, clearly showing the linear dependence at low$-z$. The net deviations from FLRW behaviour at the present day in a sample EdS universe with $h=0.45$ are $\Omega^0_\mathrm{eff}\approx 4\times 10^{-5}$ and $\left|\Delta R/R\right|\approx 3\times 10^{-5}$, while those in the WMAPV concordance model are $\Omega_\mathrm{eff}^0\approx 4.4\times 10^{-6}$ and $\left|\Delta R/R\right|\approx 4\times 10^{-6}$. For $h=0.701$, an EdS universe shows a much greater impact with an effective energy density $\Omega^0_\mathrm{eff}\approx 1.2\times 10^{-4}$, larger than that in radiation, $\Omega^0_R\approx 4.2\times 10^{-5}h^{-2}\approx 9\times 10^{-5}$. Viewed differently, this is a $0.012\%$ effect. Increasing the scalar spectral index from $n_s=0.96$ to $n_s=1.04$ increases the effective energy density for $\Lambda$CDM to $\Omega_\mathrm{eff}\approx 5.6\times 10^{-6}$, with $\left|\Delta R/R\right|\approx 5\times 10^{-6}$. The ratios between the corrections agree well with the analytic approximations and in all cases the corrections serve to decelerate the expansion.

\begin{center}
\begin{figure}
\includegraphics[width=0.45\textwidth]{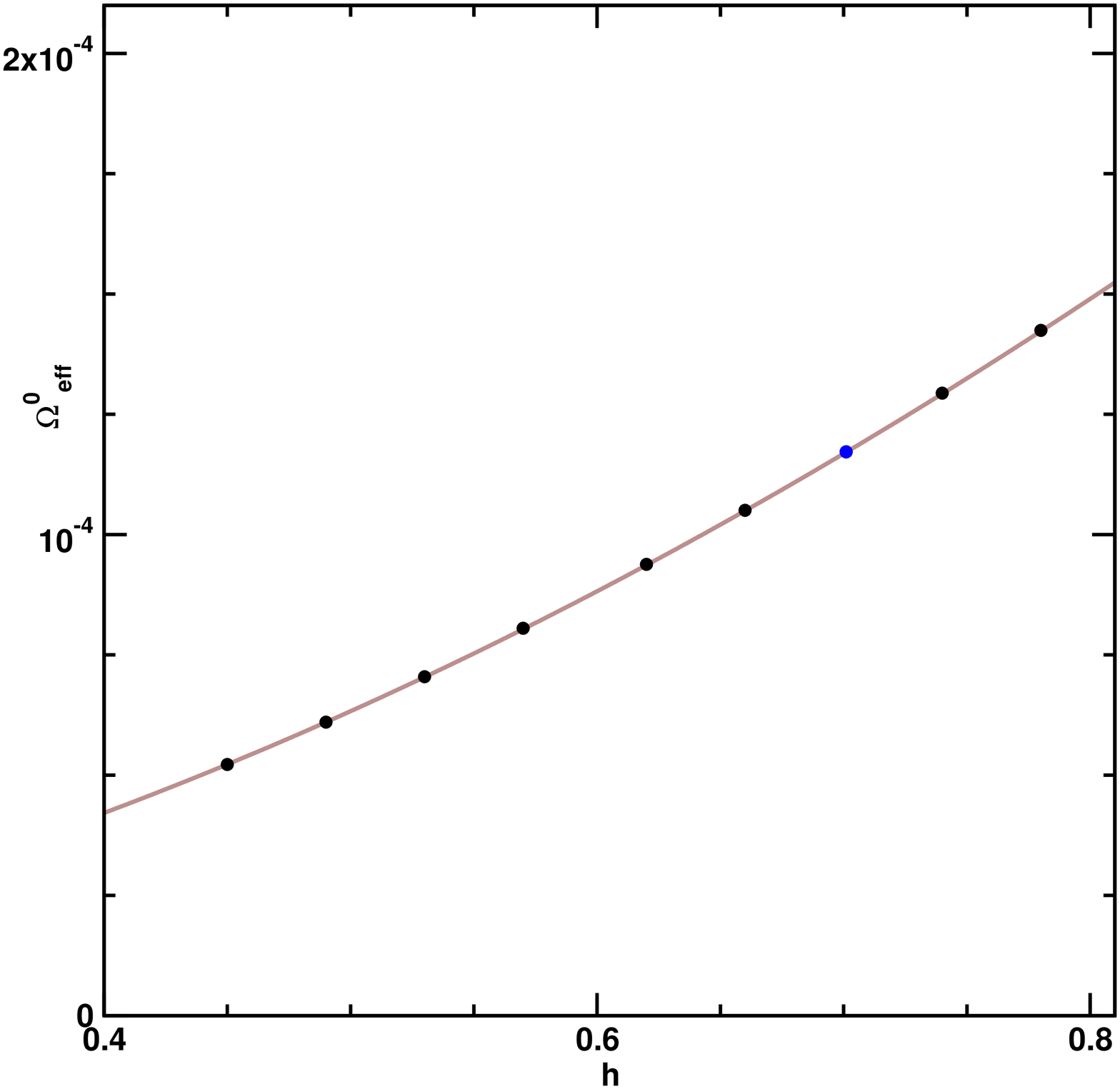}\;
\includegraphics[width=0.45\textwidth]{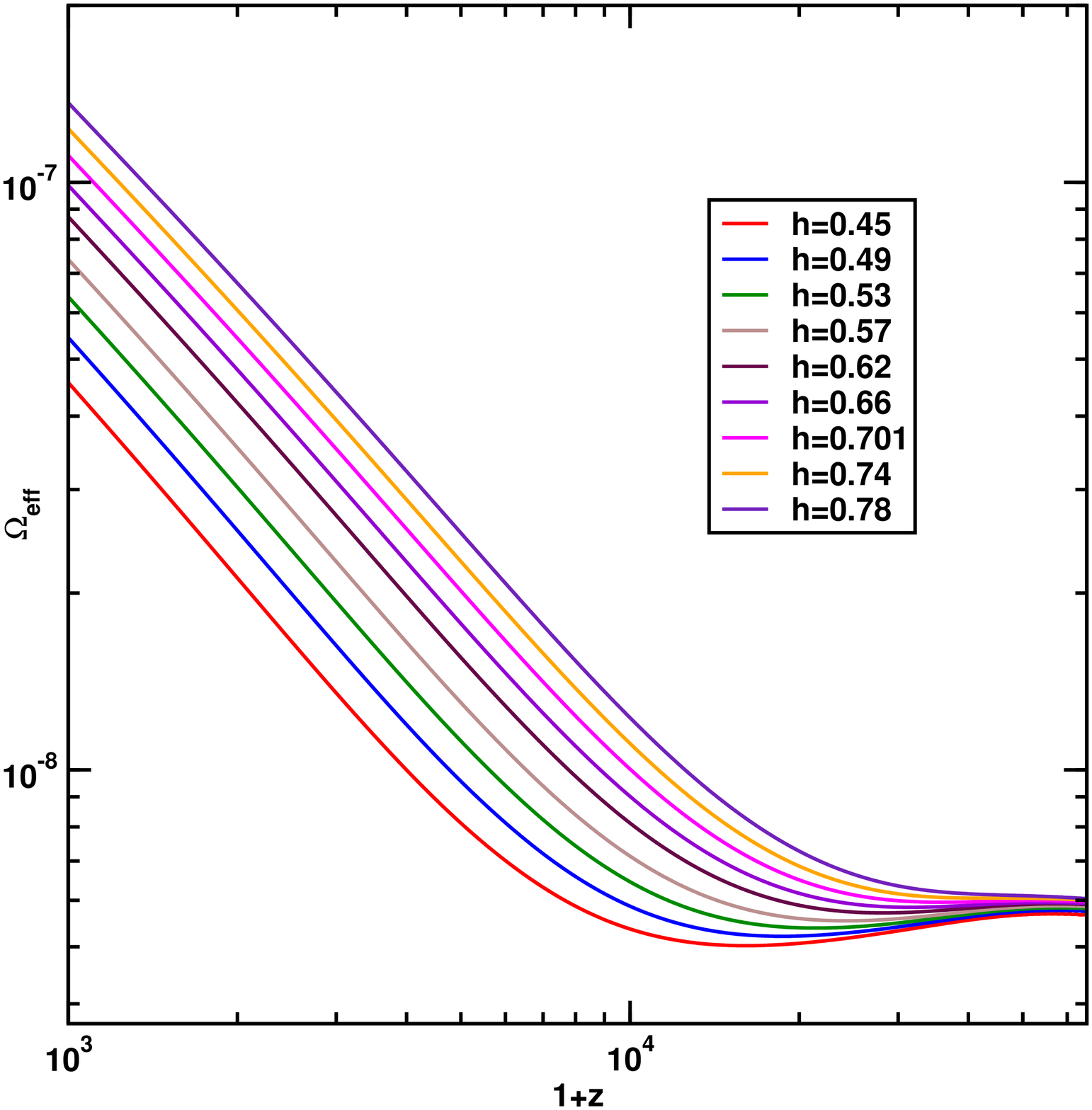}
\caption{Dependence of the backreaction on the Hubble rate for EdS models with $n_s=0.96$. Left: $\Omega^0_\mathrm{eff}$ as a function of Hubble rate. Points show the numerical results, while the curve is $\Omega_\mathrm{eff}\propto h^{1.8125}$. Right: The convergence of the modification to the Friedmann equation to a constant value deep in radiation domination.}
\label{H_w}
\end{figure}
\end{center}

Figure \ref{H_w} shows the present day effective energy density $\Omega^0_\mathrm{eff}$ for an EdS model with a range of Hubble rates and all other parameters held constant. It should be noted that these models are illustrative only, and that without adjustments to the primordial power spectrum and matter abundances they cannot fit the CMB, let alone large-scale structure. The power law
\be
  \Omega^0_\mathrm{eff}\propto h^{1.8125} \Rightarrow m=0.1875
\ee
fits our results for the range $h\in(0.4,0.8)$, which covers most values of interest.

As expected, the impact at the epoch of recombination is of the order of $10^{-8}$ and in EdS cases is larger still. Starting soon after recombination the corrections interpolate smoothly between the matter and radiation approximations presented above. The delay between recombination and the EdS behaviour is due to the photon velocities which are decaying but remain significant for a period after recombination. As predicted, the corrections tend to an approximate constant value deep in radiation domination, relatively independent of the Hubble rate. If one considers the $z>150,000$ r\'egime there is a further turnover and the backreaction begins once more to decline.

\begin{center}
\begin{figure}
\includegraphics[width=0.45\textwidth]{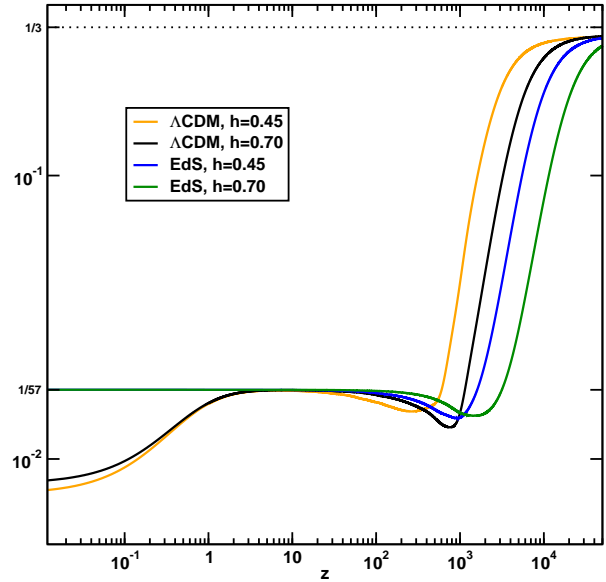}
\caption{Effective equation of state of the modifications in EdS and $\Lambda$CDM cosmologies. The asymptotic values of $w_\mathrm{eff}=1/57$ in matter domination and $w_\mathrm{eff}\approx 1/3$ are clearly visible, as is a transition at $z\approx 1000$ from the decay of radiative velocity.}
\label{weff-LCDMEdS}
\end{figure}
\end{center}

Figure \ref{weff-LCDMEdS} shows the corresponding effective equation of state for the EdS and $\Lambda$CDM models (see Figure [7] of BBR08 for comparison with the previous results). For the EdS models, $w^0_\mathrm{eff}=1/57$ as expected. For $\Lambda$CDM the transition at $z\approx 0.7$ is clearly visible, and the equation of state declines significantly as one approaches the current epoch, ending at $w^0_\mathrm{eff}\approx 1/120$ for the concordance model.
 At high redshifts, the effective equation of state rises towards one third and asymptotes at $w_\mathrm{eff}\approx 0.31$ in both EdS and $\Lambda$CDM models.

\subsubsection{Dark Energy and Quintessence}
The main dynamical alternatives to a standard $\Lambda$CDM model are quintessence cosmologies in which a scalar field, minimally-coupled to gravity, drives an acceleration of the universe at low redshift. In contrast to the $\Lambda$CDM model dynamical dark energy models contain dark energy perturbations, which contribute non-trivially to $\Td$ and $\Sd$, and alterations to $\dot{\Phi}$. As we are working on linear scales we still expect $\Qd\ll\Td$ and the dominant differences from $\Lambda$CDM to come from the dark energy perturbations. Since $\Td$ and $\Sd$ contain terms of the form $V(\av{\phi})-\av{V(\phi)}$ one might expect relatively large deviations, in models with an exponential potential for example. Indeed, in W03 deviations on the order of unity were estimated from a nonlinear, clumping model of such a field. We consider the following dark energy models:
\begin{itemize}
 \item a constant equation of state, $w_\phi=\mathrm{const}$,
 \item the standard examples for scalar field potentials: an exponential and an inverse power law,
 \item two parameterisations of the dark energy evolution, first the standard parameterisation of the equation of state,
       $w(a)=w_0+w_a(1-a)$ , and a parameterisation of the energy density which includes an early dark energy component. 
\end{itemize}
Finally, we will investigate the effect of a change of the sound speed $c_s^2$ of the dark energy component.\footnote{In this study, unless noted we take $c_s^2=1$ for a scalar field cosmology. However, the use of this in our context has recently been challenged \cite{Christopherson08}.} Unless otherwise noted, in this section we employ WMAP 5-year parameters $h=0.701$, $n_s=0.96$ and $\Omega_\phi=0.721$.

\begin{center}
\begin{figure}
\includegraphics[width=0.45\textwidth]{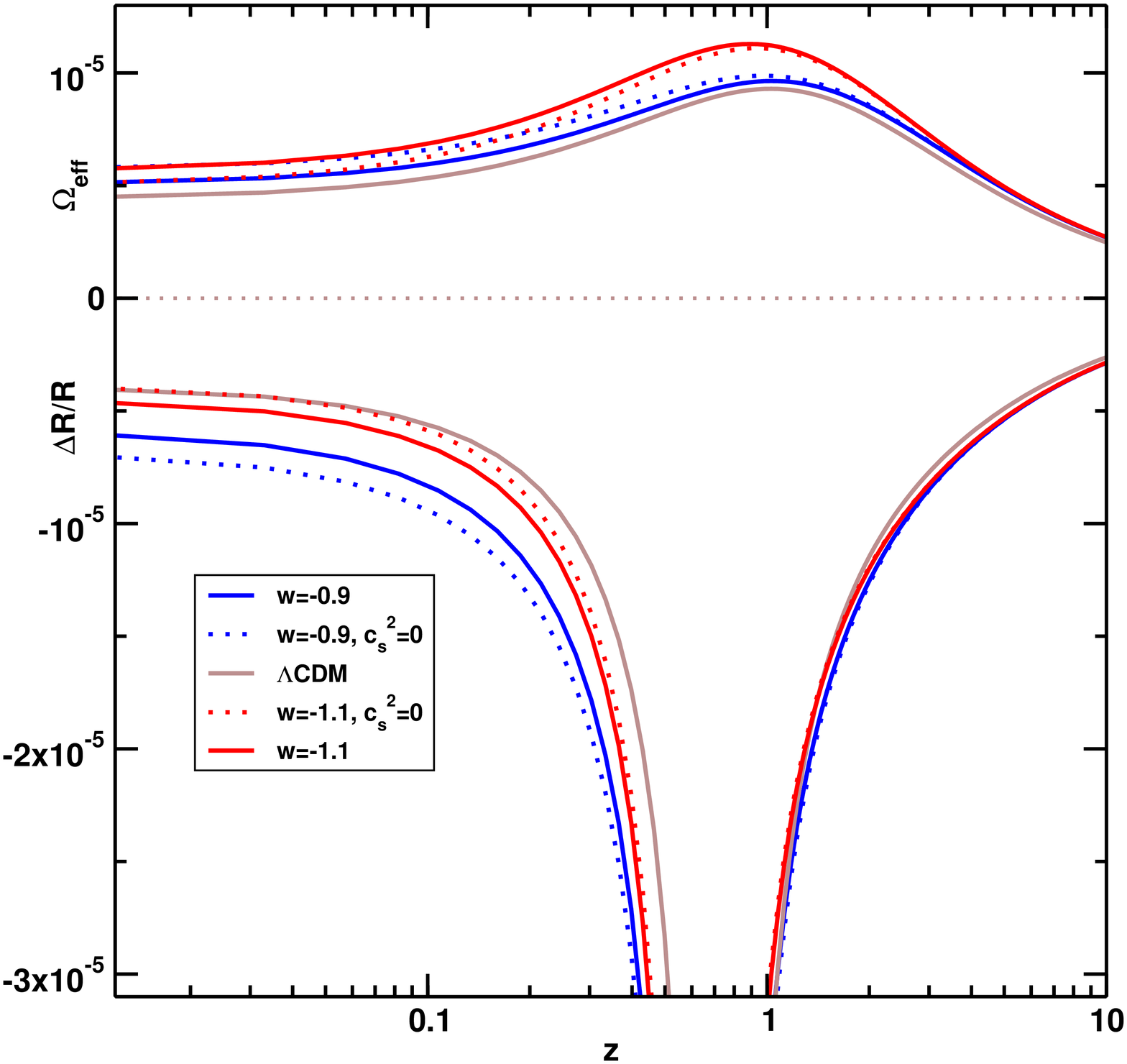}\;
\includegraphics[width=0.45\textwidth]{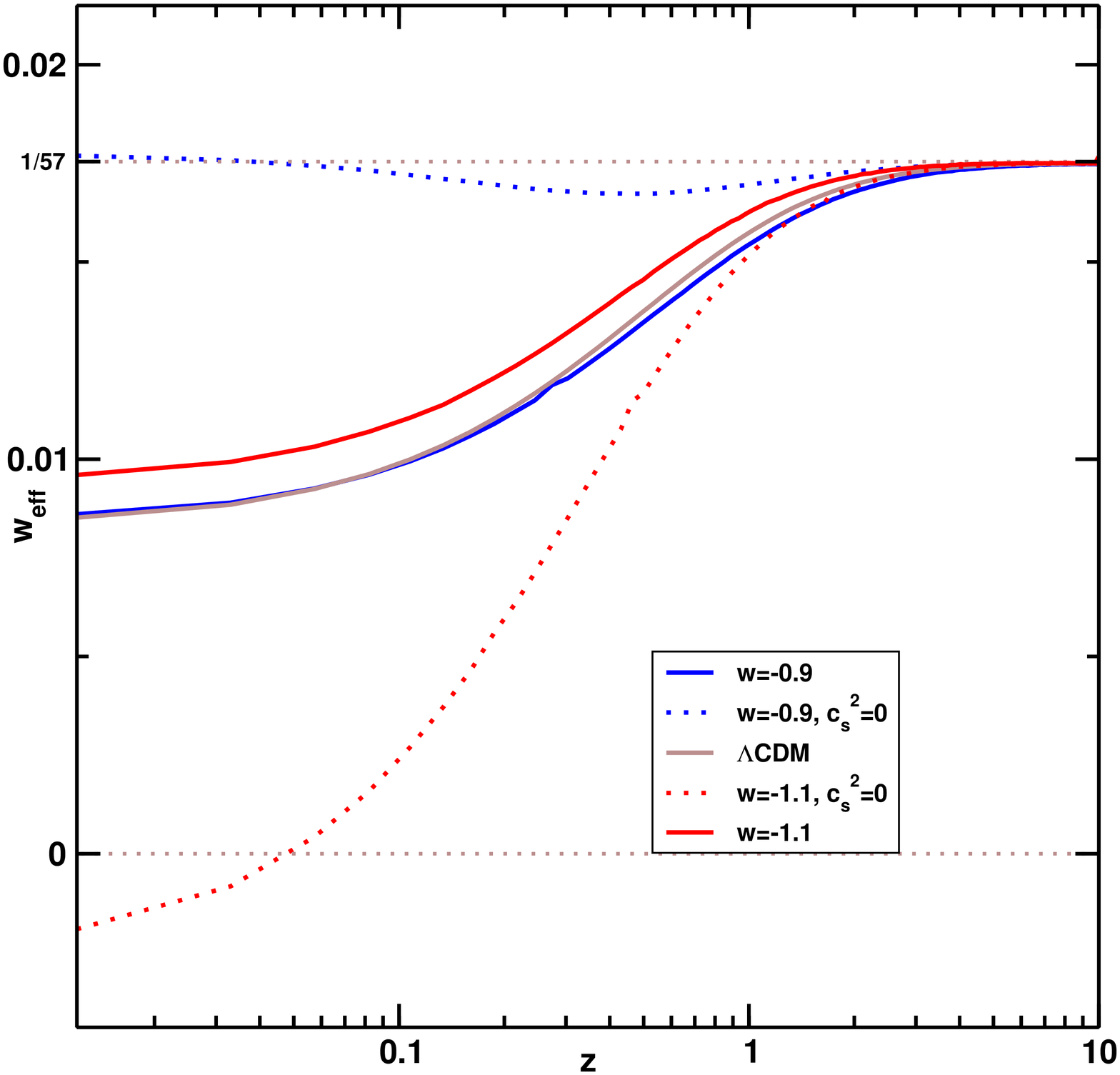}
\caption{Left: Effective energy density in backreaction and modification to the Raychaudhuri equations in $\Lambda$CDM, a quintessence model with $w_\phi=-0.9$ and a phantom model with $w=-1.1$ (solid curves). Also shown (dotted curves) are are models where the rest-frame speed of sound squared of the dark energy is set to zero. Right: The effective equations of state for these models.}
\label{ConstW}
\end{figure}
\end{center}

Dark energy models with a constant equation of state models serve as a good first example and are indistinguishable from $\Lambda$CDM for $z\gtrsim 10$. We separate models into the ``quintessence'' r\'egime with $w_\phi>-1$ and the phantom r\'egime with $w_\phi<-1$, in both cases taking $c_s^2=1$. In Figure \ref{ConstW} we plot the modifications and effective equations of state for models with $w_\phi=-0.9$ and $w_\phi=-1.1$. In both cases dark energy perturbations make the modifications larger than those in $\Lambda$CDM, with the increase in $\Sd$ making the change in $\Delta R/R$ larger than that in $\Omega_\mathrm{eff}$. The quintessence field with $w_\phi=-0.9$ in particular generates a large deviation from the standard Raychaudhuri equation. The effective equations of state in both cases are similar to $\Lambda$CDM, which occupies an envelope between the quintessence and phantom curves. For $w_\phi=-0.9$, $w_\mathrm{eff}$ is slightly less than that for $\Lambda$CDM, while for $w_\phi=-1.1$ it is slightly larger, and $\Lambda$CDM occupies an envelope between the two curves. We then expect the effective equations of state from models with $w_\phi>-1$ to be slightly less than those of $\Lambda$CDM and those from phantom fields to be greater.

\begin{center}
\begin{figure}
\includegraphics[width=0.45\textwidth]{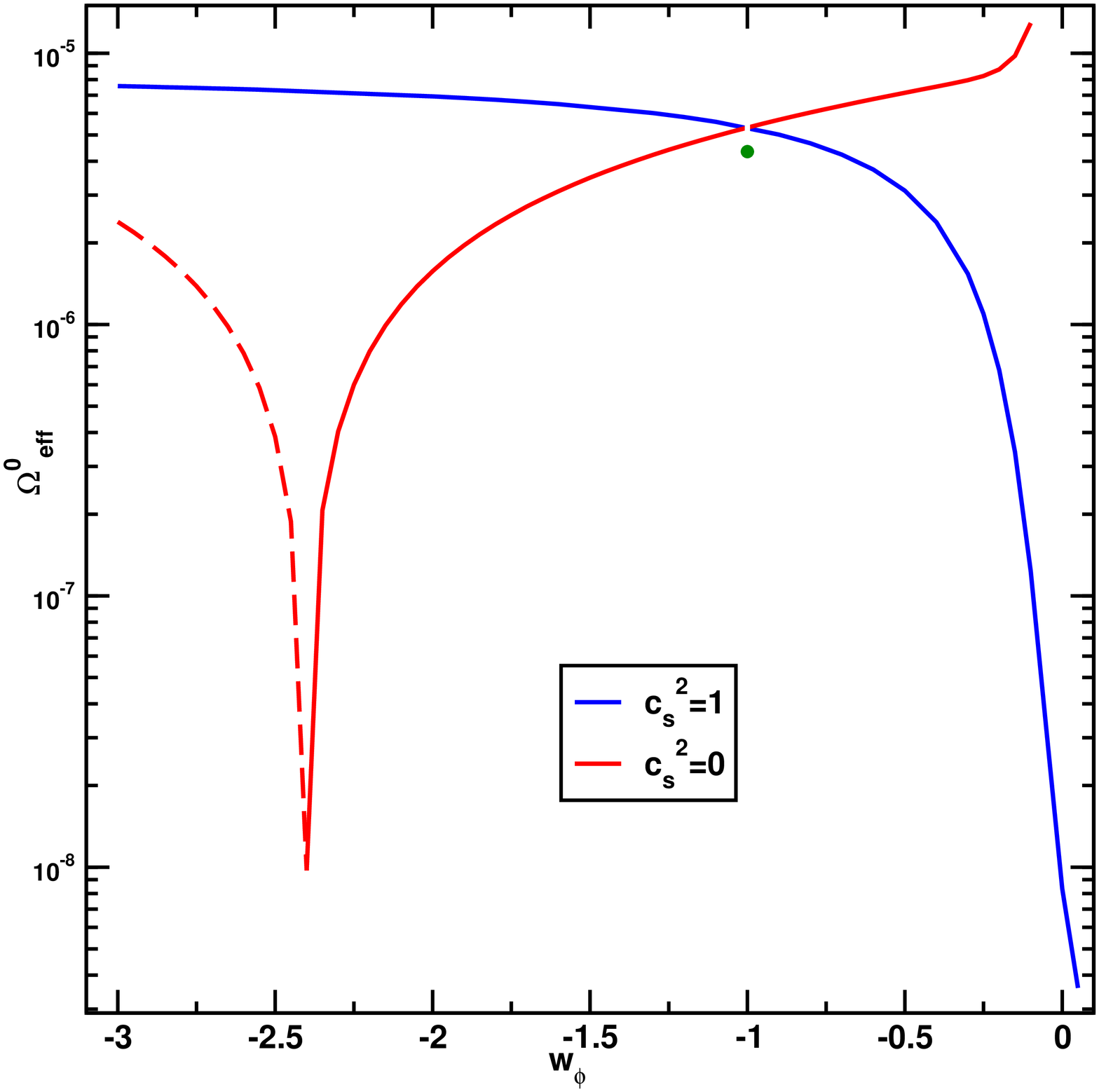}\;
\includegraphics[width=0.45\textwidth]{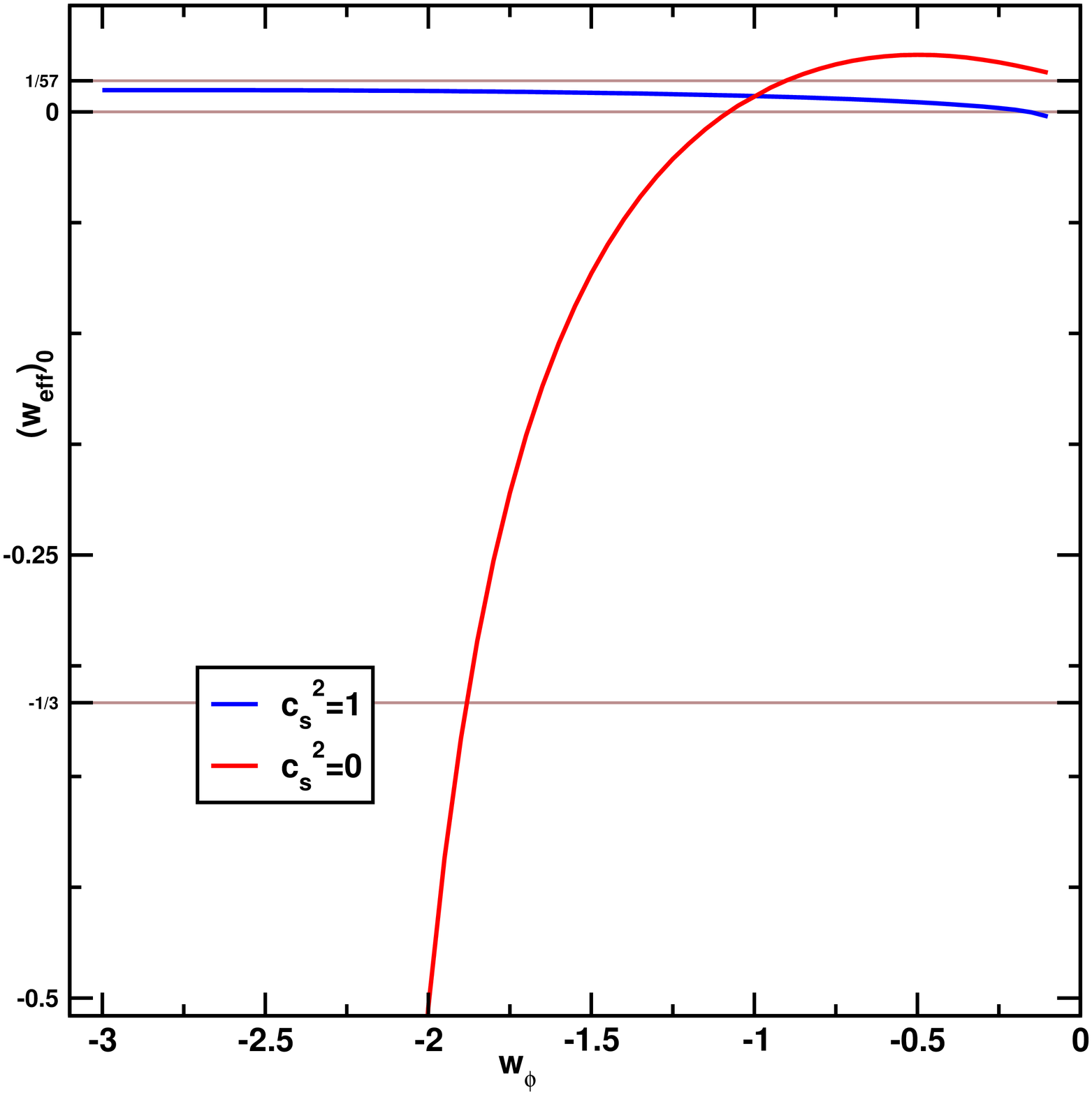}
\caption{Dark energy models with a constant equation of state and sound speed $c_s^2=1$ (blue) and $c_s^2=0$ (red). Left: $\Omega^0_\mathrm{eff}$ as a function of $w_\phi$, with the $\Lambda$CDM value marked in green. Right: Effective equation of state $w_\mathrm{eff}$ as a function of $w_\phi$.}
\label{WDep}
\end{figure}
\end{center}

The blue curves in Figure \ref{WDep} show the dependence of $\Omega^0_\mathrm{eff}$ and $w^0_\mathrm{eff}$ on $w_\phi$ for $w_\phi\in(-3,0)$. The transition from $w_\phi<-1$ to $w_\phi>-1$ is discontinuous since $\Lambda$CDM does not contain dark energy perturbations, and the $\Lambda$CDM value is plotted in green. The effective energy density in the constant-$w_\phi$ models increases monotonically in the quintessence r\'egime as $w_\phi\rightarrow -1$ but asymptotes to $\Omega^0_\mathrm{eff}\approx 8\times 10^{-6}$ in the phantom r\'egime. The effective equation of state acts as expected, with that from quintessence being greater than that from $\Lambda$CDM and that from phantom fields being less. In both cases the change is relatively slight; for the range $w_\phi\in(-3,-0.15)$ the effective equation of state is bounded by $w_\mathrm{eff}\in(0,1/57)$ and thus the modifications from a constant equation of state behave approximately as dust. For larger equations of state it becomes very negative and acts to accelerate the universe for the rather unrealistic case $w_\phi>18$.

We now turn to quintessence models with a time-varying equation of state. As a first step, we consider two examples of basic quintessence potentials, the exponential (see for example \cite{Wetterich88,Wetterich95}) and the inverse power law \cite{Ratra88}. For simplicity and to allow comparison with an EdS case, we do not break the tracking behaviour of the exponential, taking
\be
\label{eq:exponential}
  V(\phi)=C\exp(-\alpha\phi) .
\ee
We choose $C$ such that $\Omega^0_M=0.8$ and $\Omega^0_\phi=0.2$. Since this field is still tracking the matter component at the present epoch, it has an equation of state today of $w^0_\phi\approx 0$. The Ratra-Peebles model we consider instead has a potential
\be
  V(\phi)\propto 1/\phi^2
\ee
where the energy scale is set such that the energy density has the WMAPV value of $\Omega^0_\phi=0.721$, which leads to $w^0_\phi=-0.64$. This model accelerates the universe but is illustrative rather than observationally viable.

Finally, we consider two examples of dynamical dark energy models that satisfy observation constraints. Firstly, we employ the standard parameterisation of the equation of state (the `CPL' parameterisation \cite{Linder:2002et, Chevallier:2000qy}),
\be
  w(a)=w_0+(1-a)w_a,
\ee
taking $w_0=-0.9$ and $w_a=0.5$,  consistent with the results in \cite{Komatsu08}. Secondly, in order to quantify the influence of the scalar field at high redshifts we also consider an early dark energy model. This model is parameterised in terms of the energy density of the quintessence component $\od(a)$, which evolves in two distinct phases. At low redshifts, it behaves as the fractional energy density of a cosmological constant, whereas at high redshifts it contributes a constant fraction $\ode$ to the total energy density. This can then be written \cite{Doran06} as
\be
  \od(a)=\frac{\od^0-\ode\left(1-a^{-3w_0}\right)}{\od^0+\om^0a^{3w_0}}+\ode\left(1-a^{-3 w_0}\right) .
\ee
We set the equation of state today $w_0=-0.99$, and choose the fraction of dark energy at early times $\ode=0.02$, compatible with observational bounds \cite{Doran07}. The fact that the quintessence component in this parameterisation contributes a constant fraction to the total energy density at high redshifts implies an exponential potential for the scalar field at those times (with the exponent $\alpha$ in equation (\ref{eq:exponential}) changing as the Universe transitions from radiation to matter domination). For low redshifts, the scalar field potential in this model then flattens to produce the desired $w_0$.

\begin{center}
\begin{figure}
\includegraphics[width=0.45\textwidth]{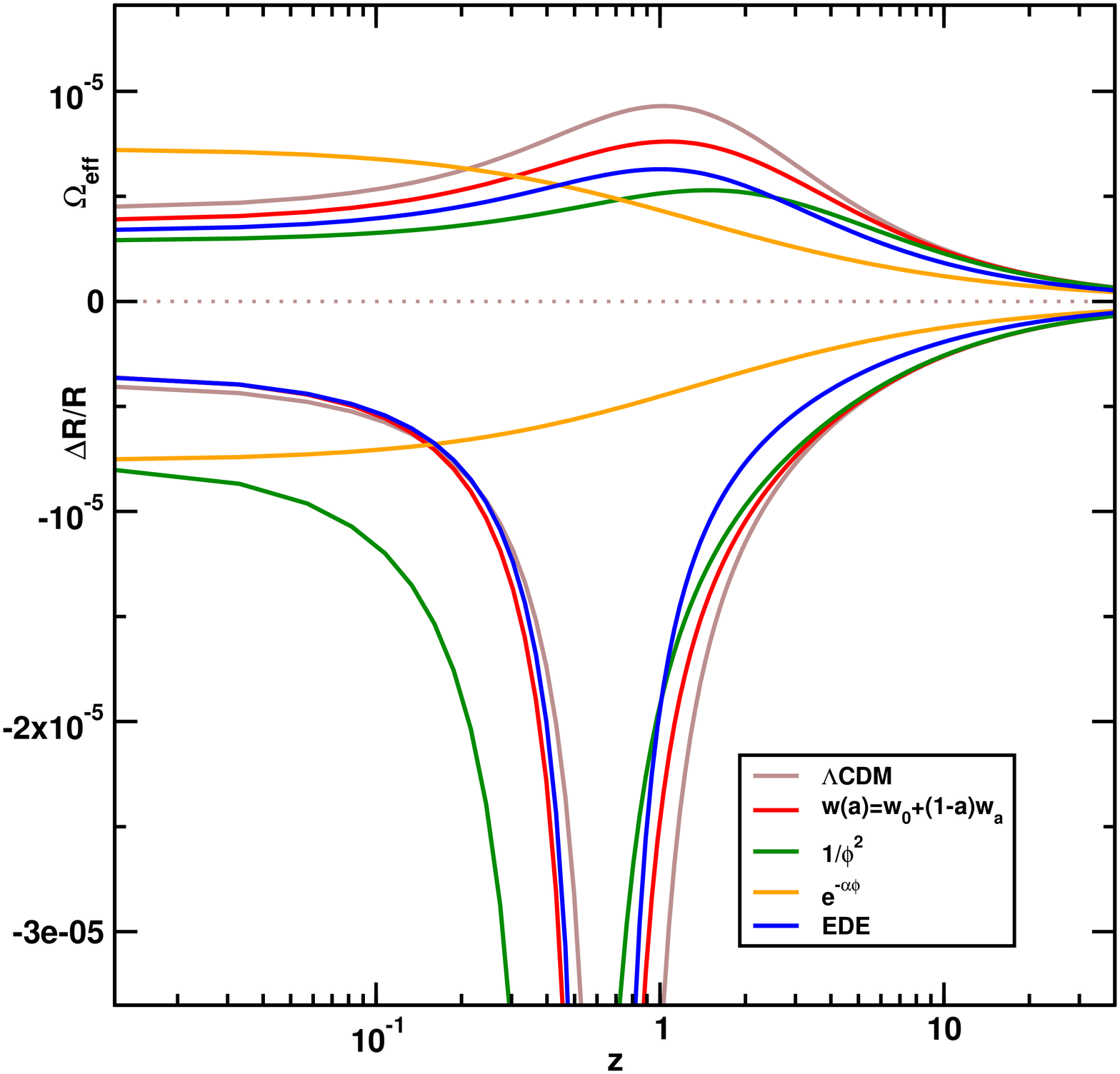}\;
\includegraphics[width=0.45\textwidth]{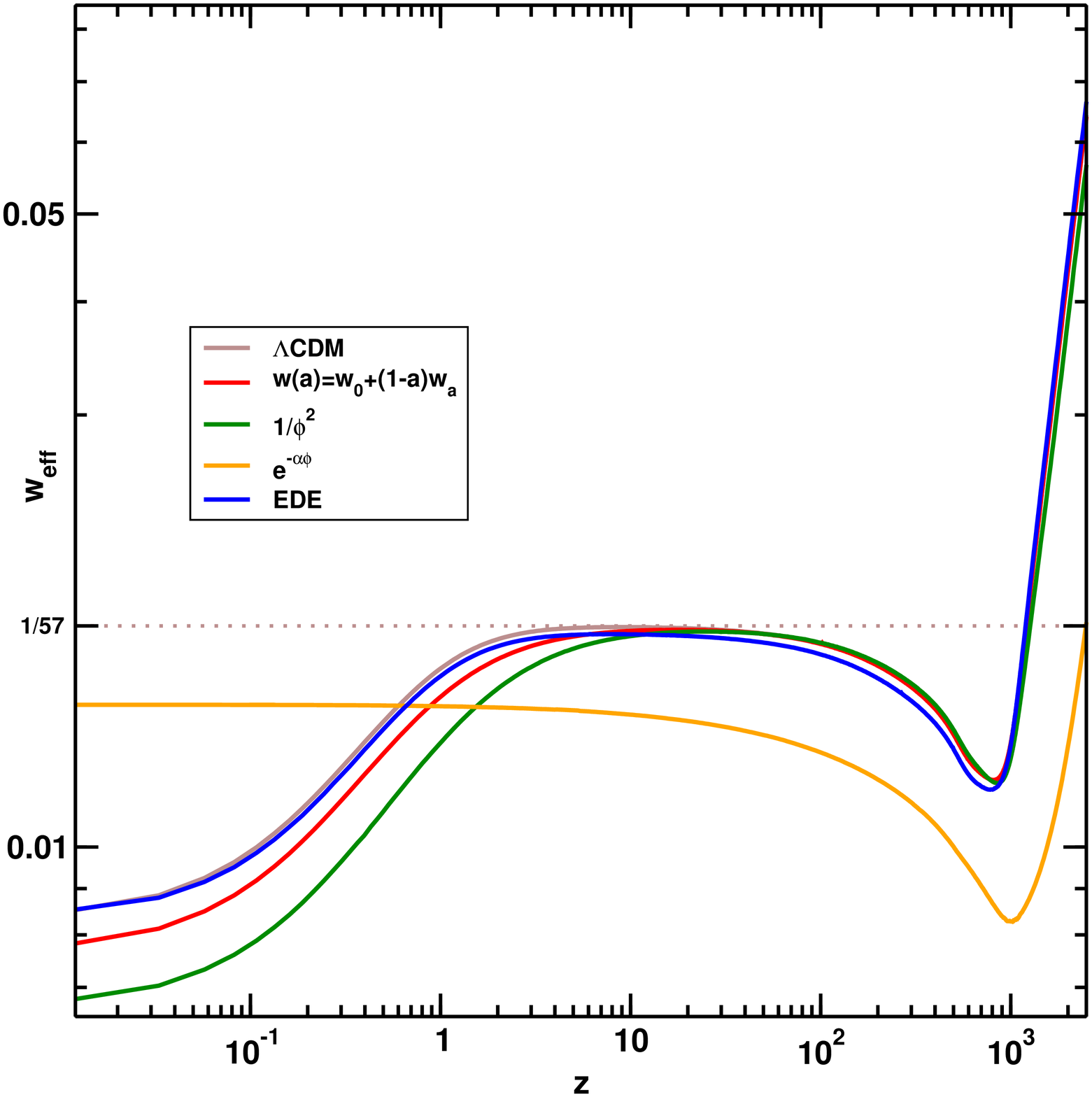}
\caption{Left: The backreaction in $\Lambda$CDM and selected dynamical dark energy models -- the linear CPL parameterisation, a Ratra-Peebles field with $V(\phi)\propto 1/\phi^2$, a tracking field with $V(\phi)\propto\exp(-\alpha\phi)$ and an early dark energy model. Right: The effective equations of state for the same models.}
\label{Q}
\end{figure}
\end{center}

In the left-hand panel of Figure \ref{Q} we plot $\Omega_\mathrm{eff}$ and $\Delta R/R$ in these dynamical models in comparison with $\Lambda$CDM. In contrast with the constant-$w_\phi$ case, the dynamical models produce an effective energy density lower than that of $\Lambda$CDM. The effective energy density at the present day in the CPL parameterisation is $\Omega^0_\mathrm{eff}\approx 3.8\times 10^{-6}$ and the impact on the Raychaudhuri equation is $\Delta R/R\approx -3.6\times 10^{-6}$. Backreaction in the model with the Ratra-Peebles field is more heavily suppressed with $\Omega^0_\mathrm{eff}\approx 2.9\times 10^{-6}$ but the deviation of the Raychaudhuri equation is significantly boosted to $\Delta R/R\approx -8.1\times 10^{-6}$. The early dark energy model produces $\Omega_\mathrm{eff}^0\approx 3.3\times 10^{-6}$, similar to and slightly larger than the Ratra-Peebles form, but generally produces small corrections to the Raychaudhuri equation. Finally, the exponential field resembles an EdS model but is heavily suppressed; as we saw earlier, an EdS model with these parameters generates $\Omega^0_\mathrm{eff}\approx 1.1\times 10^{-4}$ at the present epoch, while the exponential potential produces only $\Omega^0_\mathrm{eff}\approx 7.1\times 10^{-6}$.

The effective equations of state for these models are plotted in the right panel of Figure \ref{Q}. The CPL, Ratra-Peebles and early dark energy models all closely resemble $\Lambda$CDM until a redshift of $z\approx 10$ at which point they decline to different present-day values, with the Ratra-Peebles field the smallest at $w^0_\mathrm{eff}\approx 1/150$, the CPL parameterisation dropping to $w_\mathrm{eff}\approx 1/130$, and the early dark energy model almost indistinguishable from $\Lambda$CDM. Above $z\approx 10$ the Ratra-Peebles and CPL models are as expected equivalent to $\Lambda$CDM, and the early dark energy model is only slightly different. The differences in this case generally arise from $\Qd$, which is non-zero even in matter domination. The effective equation of state produced by the exponential potential naturally displays a very different behaviour and closely resembles that from an EdS model suppressed to $w^0_\mathrm{eff}\approx 1/70$. This is because the nature of the dark energy perturbations is very different to that of dust perturbations, and as the field is tracking matter it is not diluted relative to dust as one moves to higher redshifts.

Thus far we have only considered dark energy models based on scalar fields with a canonical kinetic term and a sound speed of $c_s^2=1$. However, many models of dark energy, such as k-essence, generically predict $c_s^2\neq 1$ \cite{Erickson:2001bq,DeDeo:2003te,Bean:2003fb,Hannestad:2005ak}. We now investigate the effects of a sound speed different from unity by returning to models with a constant equation of state but setting $c_s^2=0$, which models a dark energy allowed to cluster. The dotted lines in Figure \ref{ConstW} show $\Omega_\mathrm{eff}$ to be relatively small for both $w_\phi=-0.9$ and $w_\phi=-1.1$, while the corrections to the Raychaudhuri equation the deviations are increased further for $w_\phi=-0.9$ but suppressed to the level of $\Lambda$CDM for $w_\phi=-1.1$.\footnote{The coincidence between the dotted curve in Figure \ref{ConstW} and $\Lambda$CDM is, however, just that. Were we to take e.g. $w_\phi=-1.15$ we would observe a different impact.} The effective equation of state for $w_\phi=-0.9$ is driven back towards the EdS value. The case $w_\phi=-1.1$, however, is more interesting. While we have allowed the field to ``cluster'' -- although this interpretation is somewhat suspect for a phantom field -- the effective equation of state is negative, $w_\mathrm{eff}\approx -1/500$, although it still closely resembles dust.

The red curves in Figure \ref{WDep} show the dependence of $\Omega^0_\mathrm{eff}$ and $w^0_\mathrm{eff}$ on $w_\phi$ when $c_s^2=0$. The dependence of the effective energy density on $w_\phi$ is opposite to the standard case and coincides only at $w_\phi=-1$ where no dark energy perturbations are present. For $w_\phi>-1$ the backreaction increases to $\Omega^0_\mathrm{eff}\approx 10^{-5}$. The ``clumps'' in the field are acting as a dark matter, making this model in this respect resemble an EdS case, although the effect here is somewhat lessened. In the phantom r\'egime $\Omega^0_\mathrm{eff}$ declines smoothly but increasingly rapidly, again due to the fields dispersing structure. The effect is greatly enhanced for $c_s^2=0$ compared to $c_s^2=1$. The effective energy density becomes negative at $w_\phi\approx -2.4$ and for smaller equations of state becomes increasingly negative.

In the quintessence r\'egime the equation of state rapidly grows and we recover the EdS value $w_\mathrm{eff}\approx 1/57$ only for $w_\phi\approx -0.9$. For larger values of $w_\phi$, the effective equation of state increases to a maximum of $w_\mathrm{eff}\approx 1/30$ at $w_\phi=-1/2$ before declining again towards EdS as $w_\phi\rightarrow 0$. For the phantom case, however, $w_\mathrm{eff}$ declines rapidly with $w_\phi$ and for $w_\phi<-1.85$, we recover a backreaction which accelerates the universe, albeit with an energy density $\Omega^0_\mathrm{eff}\approx 10^{-6}$ and in a model which is already accelerating. At $w_\phi\approx -2.4$ the effective equation of state passes through negative infinity and becomes positive. However, since for these values of $w_\phi$ the impact on the Friedmann equation becomes negative, these models still act to accelerate the universe.

\section{Discussion}
\label{Discussion}
In this paper we have presented spatially averaged cosmological equations in a simple form easily applied to a wide variety of models, both numerical and analytic, for which the 3+1 split remains valid and the shift vector is vanishing. This encompasses analytic models currently considered in the literature such as LeMa\^itre-Tolman-Bondi models, swiss-cheese generalisations, the Szekeres model or other systems such as networks of McVittie metrics which could be used to generate a distribution of sources. The formalism is also applicable to purely numerical approaches, the ideal of which would be a fully relativistic $n$-body code or approximations to such. In our approach, any matter source can be expressed as an effective perfect fluid since the physics of the underlying model are left unaltered, and so we can consider any combination of fluids ranging from cosmological dust and radiation, to scalar fields, to more realistic physical fluids with non-trivial equations of state and couplings between different species.

Averaging in cosmology has in recent years typically been invoked in an attempt to solve the dark energy problem, but is entirely distinct from it and has a history dating back at least to the early 1960s. As such, we have attempted to separate the dark energy problem from the averaging problem, and considered systems including not only standard matter such as baryons and radiation, along with cold dark matter, but also included scalar fields and dark energies on an equal footing. Should scalar fields have played a part in the universe's evolution then it is vital to include them in this averaging process. While we have focused on the universe from radiation domination onwards, this statement holds true even in the inflationary universe. 

As a concrete example of the impacts one should expect, we applied our formalism to linearly-perturbed Robertson-Walker universes. While assuming perturbation theory valid restricts us to minor deviations from Robertson-Walker behaviour, the approximation is useful for two chief reasons. Firstly, until recent times the universe is extremely well described by linear theory and calculations of the backreaction from linear theory are accurate. Secondly, while the results we find for low redshifts may be incomplete they have the advantage of predictive power. A full quantitative calculation of the backreaction at recent times is currently unfeasible; as a result we are restricted either to assumptions about the nature of the universe (as in \cite{Rasanen08,Buchert08}), or to perturbation theory (as in \cite{Li08,Behrend08,Paranjape08}). Perturbation theory is well tested and understood and with modern cosmological probes many of its parameters are known to within a few percent.

We constructed modified versions of the cmbeasy and CMBFast codes to evaluate the backreactions from linear modes in a range of models. Far the greatest impacts arise in EdS models. In these models, the effective equation of state of the ``backreaction fluid'' is exactly $w_\mathrm{eff}=1/57$, while the effective energy density is dependent on the primordial power spectrum and the Hubble rate at the present day. Changes to the primordial power spectrum renormalise the impacts, while changes to the Hubble rate alter them only in matter domination. In radiation domination the deviations are independent of the Hubble rate. The effective energy density depends on the Hubble rate as $\Omega^0_\mathrm{eff}\sim h^{1.8125}$, and for a standard Hubble rate $h=0.701$ we find $\Omega^0_\mathrm{eff}\approx 10^{-4}$. This is approximately equivalent to the current energy density in radiation and it has been argued (e.g. \cite{Coley07}) that even small deviations can be significant. For a Hubble rate $h=0.45$ for which an EdS model can be forced to fit the CMB the impact is somewhat less, $\Omega_\mathrm{eff}\approx 5\times 10^{-5}$. The backreaction also evolves in matter domination with the scale factor, and so the impact at recombination is of the order of $10^{-7}$-$10^{-8}$, roughly comparable to existing CMB anisotropies. In radiation domination the impacts flatten approximately to a constant. The present-day effective energy density for the concordance $\Lambda$CDM case is $\Omega^0_\mathrm{eff}\approx 4.4\times 10^{-6}$, while the effective equation of state undergoes a transition at $z\approx 1$ and diminishes from the pure dust case to $w_\mathrm{eff}\approx 1/120$. The effective equation of state from linear modes in these two standard cases is thus always greater than zero, in contradiction to BBR08 where the pressure correction $\Sd$ was neglected. The matter-domination prediction of $w_\mathrm{eff}=1/57$ holds until close to recombination. After recombination the photon velocities decay and there is a period at which they contribute enough to $\Td$ and $\Sd$ to affect the effective equation of state, exhibited as a trough at $z\approx 1000$. Before recombination, the effective equation of state increases steadily and tends towards $w_\mathrm{eff}=1/3$.

We then considered a range of dark energy universes, taking as our test cases models with a constant equation of state $w_\phi$, the CPL parameterisation $w(a)=w_0+(1-a)w_a$, a quintessence field with an exponential potential, a quintessence field with an inverse power law potential, and a model of early dark energy. For the constant equations of state we also considered an approximation to a clumping scalar field by setting the field's rest-frame speed of sound to zero.

In the region of $w_\phi=\mathrm{const}\approx -1$ with $c_s^2=1$, the dark energy generically creates a slightly larger effective energy density and more significant deviations from the usual Raychaudhuri equation than $\Lambda$CDM. The equation of state is slightly lower for a quintessence field with $w_\phi>-1$ and slightly higher for a phantom with $w_\phi<-1$. Allowing $w_\phi$ to vary more widely, the effective energy density tends to vanish as $w_\phi\rightarrow 0$ where $w_\mathrm{eff}$ is marginally negative. For realistic models, then, the backreaction remains significantly smaller than the EdS case, and behaves as a dark matter with $w_\mathrm{eff}>0$. For strongly phantom cases with $w_\phi\rightarrow -3$ the effective energy density asymptotes to $\Omega^0_\mathrm{eff}\approx 8\times 10^{-6}$ and $w_\mathrm{eff}\approx 1/57$.

When considering dynamical models we set our parameters from WMAPV and as a result selected cases which closely resemble $\Lambda$CDM at the background level, the exception being the exponential field which resembles EdS. In all cases the backreactions are relatively small. For the CPL parameterisation we recovered backreactions similar in both form and magnitude to the $\Lambda$CDM case, although with a slightly smaller present-day effective energy density and equation of state. While its effective energy density is very close to $\Lambda$CDM, the inverse power law model produced a large correction to the Raychaudhuri equation and an effective equation of state of $w_\mathrm{eff}\approx 1/130$. The early dark energy model was tuned to closely resemble $\Lambda$CDM at low redshifts, with $w_0=-0.99$ and a dark energy density at early times of $\Omega^e_d=0.02$; unsurprisingly, the backreactions at low redshifts are very close to $\Lambda$CDM although again the effective energy density is marginally reduced. At earlier times, the effective equation of state diverges slightly from the $\Lambda$CDM case due to a non-vanishing $\Qd$ even in matter domination, although this discrepancy disappears before recombination when photon-baryon coupling alters the perturbations significantly.

Finally, we chose to keep the exponential field as a tracker, to more closely resemble EdS. The backreaction produced in this model is surprisingly small, with $\Omega^0_\mathrm{eff}\approx 7\times 10^{-6}$, and the equation of state is reduced to approximately $w_\mathrm{eff}=1/70$ until recombination at which point it again tends towards $w_\mathrm{eff}=1/3$. The reduced effective equation of state stems from the different behaviour of the dark energy perturbations; even though the equation of state of the field is dustlike, the perturbations differ from dust perturbations. Since the field is tracking matter, $\Omega_\phi/\Omega_M$ does not decay for increasing redshift and $w_\mathrm{eff}$ is thus suppressed at all times.

In all these cases the effective equation of state is insufficiently negative to accelerate the universe and the backreactions act as dust. This implies that the backreaction on sub-horizon scales acts as a brake on the universe's acceleration and a suitably-chosen smaller-scale model might generate a significant impact. This would be similar to \cite{Li01} where the backreaction from super-horizon modes is used to stop the quintessential expansion.

As a final case we took fields with a constant equation of state and a sound speed $c_s^2=0$, changing their behaviour dramatically. With $w_\phi=0$ the backreaction is relatively large, $\Omega^0_\mathrm{eff}\approx 10^{-5}$. As this is an approximation of a clumping field, the similarity to EdS is perhaps not surprising. Reducing $w_\phi$ towards $w_\phi=-1$ rapidly decreases the effective energy density of the backreaction. For fields with $w_\phi>-1$ the effective equation of state reaches a peak of $w_\mathrm{eff}\approx 1/30$ for $w_\phi\approx -1/2$, with $w_\mathrm{eff}\approx 1/57$ at $w_\phi\approx -0.05$ and $w_\phi\approx -0.9$. In the phantom r\'egime, the effective energy density of the backreaction plummets. For $w_\phi<-1.85$, $w_\mathrm{eff}<-1/3$ and so the backreaction acts to further accelerate the universe. For $w_\phi<-2.4$ the effective energy density becomes negative and the equation of state positive. Backreaction can then serve to generate a volume-averaged expansion which accelerates, even at linear order, albeit at a very low level and in an otherwise unappealing model. It is worth emphasising that while the behaviour with $c_s^2=0$ approximates a nonlinear behaviour the magnitude of the effect is a purely linear result. Were we to consider a smaller-scale model we might expect a larger backreaction. This may also suggest that for less pathological cases with $w_\phi>-1.85$, the backreaction could act as a brake on the phantom somewhat.

When combined with the results of W03, our study of quintessence cosmologies suggests that further analysis of scalar field cosmologies on smaller scales is required. As with any gravitating source, scalar fields are not entirely homogeneous. At linear scales we have found that these perturbations generate an effective fluid with $w_\mathrm{eff}>0$ for standard models: re-averaged perturbations in quintessence models do not act to accelerate the universe. If we additionally set $c_s^2=0$ to approximate a clumping field, $w_\mathrm{eff}$ is driven even further from zero, although as the calculation is linear the amplitude remains small. The approximate non-linear model of a clumped cosmon in W03 allowed a better approximation of the amplitude, giving $\Omega^0_\mathrm{eff}\approx\mathcal{O}(1)$ and $w_\mathrm{eff}\approx -1/15$. Since perturbed quintessence models generate such high equations of state, and allowing the field to clump alters the behaviour still further, we must confirm that an inhomogeneous field behaves on average as a homogeneous field. In forthcoming work we intend to tackle this issue employing a direct spatial averaging approach.

Our approach is not without its issues. The most obvious of these is that due to our methods we have been restricted to spatially-flat universes. This we are addressing in forthcoming work, again through the use of direct spatial averages. More fundamental issues concern both our assumption that the shift vector vanishes and, more fundamentally yet, our averaging procedure. The spatial averaging approach averages only scalar projections of the Einstein tensor, and these do not contain all the dynamics of the inhomogeneous manifold. It also intrinsically requires a 3+1 split, while a more general procedure -- Zalaletdinov's macroscopic gravity, for example -- can instead perform averages in four-dimensional domains. Most directly, we can consider extensions that employ the current formalism more generally, the obvious case being nonlinear perturbation theory. While again we would expect only minor deviations from standard Robertson-Walker behaviour, this would let us control the scaling of the velocity correctly and generalise the nonlinear approximations in BBR08 to non-EdS and non-concordance models.

In summary, we have presented a general approach to spatial averaging in cosmology applicable to multifluids in a wide variety of metrics, and demonstrated its use with a range of linearly-perturbed Robertson-Walker models. This raises interesting questions about the nature of scalar fields, which should be addressed in the near future. We have constructed codes which can consider quantitatively any linear Robertson-Walker model. For a model including a strongly phantom field with a vanishing speed of sound a backreaction with an equation of state $w_\mathrm{eff}<-1/3$ is recovered, albeit with a low effective energy density. For standard models the backreaction at linear scales remains small and with an equation of state approaching dust.

\begin{acknowledgments}
The authors would like to thank Christof Wetterich, Roy Maartens, Karim Malik, Thomas Buchert, Dominik Schwarz, Syksy R\"as\"anen, Lily Schrempp, Robert Brandenberger, Alan Coley and an anonymous referee for useful discussions and comments. IB acknowledges the Heidelberg Graduate School for Fundamental Physics for financial support. GR acknowledges support by the Deutsche Forschungsgemeinschaft, grant TRR33 `The Dark Universe'. This work was supported in part by Perimeter Institute for Theoretical Physics.
\end{acknowledgments}

\bibliography{AveragingRobertsonWalkerCosmologies}

\end{document}